\documentclass[structabstract]{aa}

\usepackage{graphicx}
\usepackage{txfonts}
\usepackage{natbib}
\bibpunct{(}{)}{;}{a}{}{,}
\usepackage{array}
\usepackage{xspace}
\usepackage{lscape}
\usepackage{url}

\newcommand{\fcov}{{\it fcov}\xspace}
\newcommand{\FeKa}{Fe K\ensuremath{\alpha}\xspace}
\newcommand{\kms}{\ensuremath{\mathrm{km\ s^{-1}}}\xspace}
\newcommand{\NH}{\ensuremath{N_{\mathrm{H}}}\xspace}
\newcommand{\redchi}{\ensuremath{\chi _\nu ^2}\xspace}
\newcommand{\slab}{{\it slab}\xspace}
\newcommand{\xabs}{{\it xabs}\xspace}
\newcommand{\xmm}{{\it XMM-Newton}\xspace}
\newcommand{\chandra}{{\it Chandra}\xspace}

\begin{document}
   
\title{The warm absorber and X-ray variability of the Seyfert 1 galaxy NGC 3516 as seen by the XMM-Newton RGS}

\author{M. Mehdipour\thanks{\email{missagh.mehdipour@ucl.ac.uk}} \and G. Branduardi-Raymont \and M. J. Page}

\offprints{M. Mehdipour}

\institute{Mullard Space Science Laboratory, University College London, Holmbury St. Mary, Dorking, Surrey, RH5 6NT, United Kingdom}

\date{Received August 2009 / Accepted February 2010}

\abstract{}{}{}{}{}
\abstract
{}
{We present a new analysis of the soft and medium energy X-ray spectrum of the Seyfert 1 galaxy NGC 3516 taken with the Reflection Grating Spectrometer (RGS) and European Photon Imaging Camera (EPIC) on board the \xmm observatory. We examine four observations made in October 2006. We investigate whether the observed variability is due to absorption by the warm absorber and/or is intrinsic to the source emission.}
{We analyse in detail the EPIC-pn and RGS spectra of each observation separately.}
{The warm absorber in NGC 3516 is found to consist of three phases of ionisation, two of which have outflow velocities of more than 1000 \kms. The third phase (the least ionised one) is much slower at 100 \kms. One of the high ionisation phases, with $\log \xi $ of 2.4, is found to have a partial covering fraction of about 60\%. It has previously been suggested that the passage of a cloud, part of a disk wind, in front of the source (producing a change in the covering fraction) was the cause of a significant dip in the lightcurve during one of the observations. From our modelling of the EPIC-pn and RGS spectra, we find that variation in the covering fraction cannot be solely responsible for this. We show that intrinsic change in the source continuum plays a much more significant role in explaining the observed flux and spectral variability than originally thought.}
{}


\authorrunning{M. Mehdipour et al.}
\titlerunning{XMM-Newton high resolution X-ray spectroscopy of NGC 3516}
\maketitle
%
\section{Introduction}
Recent X-ray observations of many Seyfert 1 galaxies and some quasars show evidence of absorbing photoionised gas in our line of sight towards the active nucleus \citep[e.g.][]{Kaa00, Ree03, Blu05, Cos07, Kaa08}. This gas is normally found to be outflowing with often several distinct components, and is referred to as a ``warm absorber"; its absorption signatures are clearly seen in the high resolution X-ray spectra. AGN outflows, in general, are believed to have important astrophysical implications. For example, strong outflows from the AGN can affect the growth of the supermassive black hole at its core and enrichment of the ISM/IGM could alter the course of the host galaxy evolution through feedback processes. The extent of the outflow contribution depends on its kinetic luminosity and the mass outflow rate. To reconstruct the kinetic luminosity of the outflow, the ionisation state and structure of the warm absorber need be to determined from photoionisation modelling of the high resolution X-ray spectra.

The first warm absorber identification was reported by \citet{Hal84} from an observation of the \object{QSO MR2251-178} with the {\it Einstein Observatory}. Then in the days of {\it ROSAT} and {\it ASCA}, warm absorbers were found through the detection of what was interpreted as broad and blended continuum absorption edges of ions such as \ion{O}{vii} and \ion{O}{viii}. For examples of early studies of AGN warm absorbers using {\it ROSAT} see \citet{Nan92}, \citet{Nan93} and \citet{Tur93}; for those using {\it ASCA}, see \citet{Rey97} and \citet{Geo98}. However, the instruments used at that time had insufficient spectral resolution to measure any details, such as the presence and parameters of narrow absorption lines. With the advent of high resolution X-ray spectrometers onboard \xmm and {\it Chandra}, the availability of high-resolution spectra has allowed a major leap forward in the study of AGN outflows in the last decade, starting with the original discovery of numerous blue-shifted absorption lines from photoionised, outflowing gas in \object{NGC 5548} by \citet{Kaa00}, the subsequent identification of similar features in a number of AGN and the discovery of an unresolved transition array (UTA) of low ionised iron ions in \object{IRAS 13349+2438} by \citet{Beh01}.

\object{NGC 3516} is a Seyfert 1.5 SB0 galaxy at a redshift of $0.008836$ \citep{Kee96} in the constellation of Ursa Major. Even from early X-ray observations of this object, signatures of a multi-phase warm absorber have been evident. From simultaneous far-UV and {\it ASCA} X-ray observations in 1995, a warm absorber with at least two absorption components was reported by \citet{Kri96}. The two components differ by a factor of 8 in ionisation parameter; the more highly ionised has a column density twice as large as the less ionised component. \citet{Net02} investigated spectral variations of NGC 3516 over a period of 7 years, by using archival {\it ASCA} and early {\it Chandra} observations. They reported a large drop in flux (factor of $\sim 50$ at 1 keV) between an {\it ASCA} observation in 1994 and the {\it Chandra} observation in 2000. \citet{Net02} concluded that the variations in the observed flux and spectra at these epochs were consistent with a constant column density of line-of-sight material reacting to changes in the ionising continuum.

NGC 3516 was observed twice by \xmm in April and November 2001; both observations were partially overlapping with {\it Chandra} observations. \citet{Tur02} presented results from the simultaneous {\it Chandra} High Energy Transmission Grating (HETG) and \xmm observations made in November 2001: analysis of the \FeKa regime showed evidence of several narrow emission features and of rapid evolution of the \FeKa line during the observation. From the 2001 observations, \citet{Tur05} reported the presence of three distinct zones (phases) of gas with different ionisation parameters covering the active nucleus. The ionisation parameter ($\xi$) is defined as
\begin{equation}
\xi  = \frac{L_\mathrm{ion}}{{nr^2 }}
\label{ion_eq}
\end{equation}
where $L_\mathrm{ion}$ is the luminosity of the ionising source over the 1--1000 Ryd band (in $\mathrm{erg}\ \mathrm{s}^{-1}$), $n$ the hydrogen number density (in $\mathrm{cm}^{-3}$) and $r$ the distance between the ionised gas and the ionising source (in $\mathrm{cm}$). Therefore, $\xi$ is in units of $\mathrm{erg\ cm\ }\mathrm{s}^{-1}$.

The warm absorber phases found in \citet{Tur05} are: a low-ionisation UV/X-ray absorber with $\log\ \xi \sim -0.5$ and hydrogen column density of $\NH \sim 0.5 \times 10^{22}\ \mathrm{cm}^{-2}$; a more highly ionised gas with $\log\ \xi \sim 3.0$ and $\NH \sim 2 \times 10^{22}\ \mathrm{cm}^{-2}$ outflowing at a velocity of $\sim 1100\ \kms$; and a phase with $\log\ \xi \sim 2.5$ and a very large hydrogen column density of $\NH \sim 25 \times 10^{22}\ \mathrm{cm}^{-2}$ covering $\sim 50\%$ of the continuum. \citet{Tur05} found the spectral variability in the 2001 observations to be consistent with the ionisation-state of the absorbing gas layers responding to the continuum flux variation.

More recently, four observations of NGC 3516 were performed by \xmm in October 2006, interwoven with {\it Chandra} observations. The analysis of the EPIC-pn and {\it Chandra} HETG data is published in \citet{Tur08}. They discovered a previously unknown ionisation phase in the \FeKa regime, with $\log\ \xi \sim 4.3$ and $\NH \sim 26 \times 10^{22}\ \mathrm{cm}^{-2}$. In line with the results from the 2001 observations, the presence of a phase with a covering fraction of $\sim 50\%$ was confirmed. This phase was found to have $\log\ \xi \sim 2.2$ and $\NH \sim 20 \times 10^{22}\ \mathrm{cm}^{-2}$. The source showed significant flux variability between observations, especially a dip in the lightcurve of the third observation. \citet{Tur08} concluded that changes in the covering fraction of the phase partially covering the continuum provide a simple explanation of the dip in the lightcurve. They interpreted this as an eclipse of the continuum due to passage of a cloud across the line of sight over half a day.

In this paper, we present a new analysis of the EPIC-pn, RGS and OM data from the four \xmm observations of October 2006, with the emphasis put on the RGS spectra. Section 2 describes the observations and data analysis and Sect. 3 focuses on the lightcurves; the spectral modelling is described in detail in Sects. 4 and 5; we discuss our findings in Sect. 6 and give concluding remarks in Sect. 7.


\section{Observations and data analysis}
We extracted the EPIC-pn, RGS and OM data for the four 2006 observations from the \xmm Science Archive (XSA).
The \xmm observation IDs are 04012010401 (51.6 ks), 04012010501 (68.8 ks), 04012010601 (68.2 ks) and 04012011001 (67.8 ks). Hereafter, we refer to them as Obs. 1, Obs. 2, Obs. 3 and Obs. 4, respectively.

All the Observation Data File (ODF) data were processed using SAS v7.12, unless specified otherwise in the text. For EPIC-pn, RGS1 and RGS2, periods of high flaring background were filtered out before extracting scientific products. For the EPIC-pn, this was done by creating Good Time Intervals (GTIs), and using them in conjunction with the internal GTI tables to remove periods with single event count rates exceeding 0.4 counts per second in the energy range $E > 10$ keV. For the RGS instrument, in addition to the internal GTIs, we constructed an extra GTI table to filter out periods with count rates above 0.4 counts per second in CCD number 9, which is the closest to the optical axis of the telescope and therefore the most affected by background flares.

The spectra and lightcurves of EPIC-pn, operating in the Small Window mode, were extracted from a circular region of $40''$ radius centred on the source. The background was extracted from a source-free region of the same radius, located at the same RAWY as the source on the same chip. The EPIC-pn showed no evidence of pile-up. The EPIC-pn lightcurves were background-subtracted and corrected (using the {\tt epiclccorr} SAS v8 task) for various effects on the detection efficiency such as vignetting, bad pixels, chip gaps, Point Spread Function (PSF) variation and quantum efficiency; the task also makes corrections which vary with time affecting the stability of the detection, such as dead time and GTIs. The RGS instruments were operated in the standard Spectro+Q mode. The first and second order spectra of RGS1 and RGS2 were extracted, and the response matrices were generated using the {\tt rgsproc} processing task.

The OM was operated in Image+Fast mode in all observations, as well as in Science User Defined mode in the last two exposures of Obs. 1, Obs. 3 and Obs. 4. The images in all four observations were taken using the U filter (3000--3900 $\AA$). The OM Image mode data were processed using the $\mathtt{omichain}$ pipeline. The resulting images are each of 1400 s exposure. The FWHM of the source is about 1.6$\arcsec$ which is consistent with the OM on-board PSF FWHM of 1.55$\arcsec$ in the {\it U} filter. We performed photometry on each image in a fully interactive way using the $\mathtt{omsource}$ program: we carefully selected the source and background regions to extract the count rate, and applied all the necessary corrections such as for the PSF and coincidence losses, including time-dependent sensitivity (TDS) corrections. The OM lightcurves were extracted from a circle of 8 pixels radius (1 pixel $=$ 0.48\arcsec) centred on the source nucleus. The background was extracted from a source-free region of the same radius.

\begin{figure}[!]
\centering
\resizebox{\hsize}{!}{\includegraphics[angle=0]{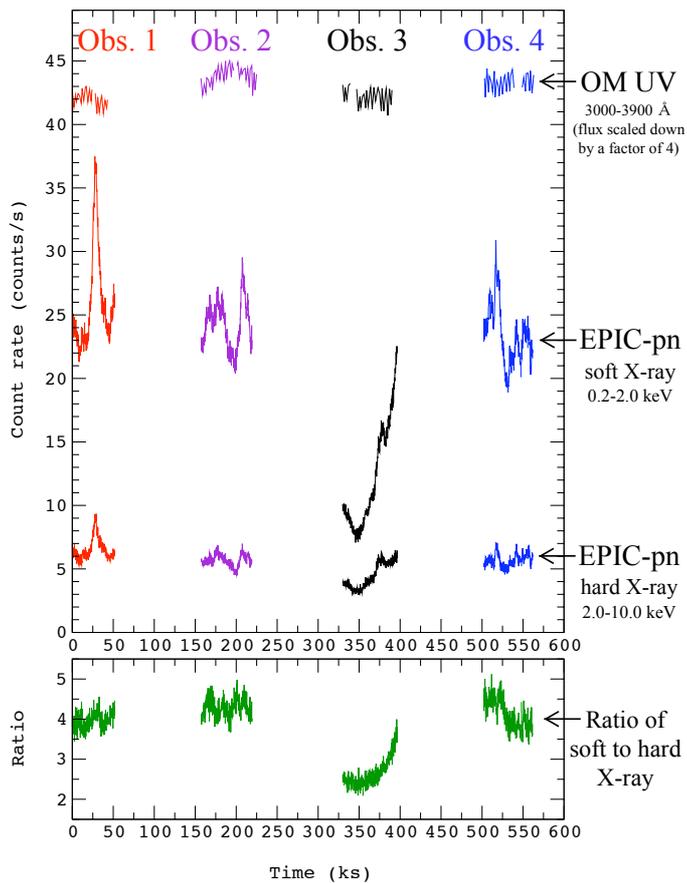}}
\caption{Upper panel: {\it Top lightcurves}: the OM background-subtracted {\it U} band (3000--3900 $\AA$) lightcurves of Obs. 1 (red), Obs. 2 (purple), Obs. 3 (black) and Obs. 4 (blue); each count rate is scaled down by a factor of 4 for clarity of display and corresponds to an OM exposure of 1400 s. {\it Middle lightcurves}: the EPIC-pn background-subtracted lightcurves of the four observations in 200 s time bins over 0.2--2.0 keV. {\it Bottom lightcurves}: the EPIC-pn background-subtracted lightcurves of the four observations in 200 s time bins over 2.0--10.0 keV. Lower panel (green): the ratio of 0.2--2.0 keV (soft X-ray) to 2.0--10.0 keV (hard X-ray) lightcurves.}
\label{PN_lightcurve}
\end{figure}

All the results presented in this work are from simultaneous fits of the EPIC-pn and the first-order RGS1 and RGS2 spectra, unless specified otherwise in the text. The EPIC-pn spectra are in the range 0.2--10 keV and are binned to a minimum of 200 counts per bin. The RGS data are in the range 6--38 $\AA$ (0.33--2.07 keV), binned by a factor of 3 (this corresponds to an average bin size of 0.04 $\AA$, which still over-samples the RGS resolution element of $\sim$ 0.07 $\AA$ FWHM). All the spectra shown in this paper are background-subtracted and are in the observed frame. The spectral analysis and modelling was done using the SPEX (version 2.01.02) fitting code \citep{Kaa96}. All the parameter errors quoted in this paper correspond to a $\Delta \chi^{2}$ of 2.

%

\section{X-ray and UV lightcurves}

Figure \ref{PN_lightcurve} shows the OM (UV, top) and EPIC-pn (X-ray) lightcurves of the four observations. The source shows significant X-ray flux variability between observations, notably the flare in Obs. 1 and the dip in Obs. 3. From visual inspection of the X-ray lightcurves, the variations in the 0.2--2.0 keV (soft X-ray) and 2.0--10.0 keV (hard X-ray) bands follow each other closely; however in Obs. 3 the ratio of soft to hard X-ray count rates clearly increases as the observation progresses indicating spectral variability; some evidence for a softer spectrum when brighter is also present in Obs. 1 and Obs. 4.

In contrast to the X-ray, the UV lightcurve shows no significant flaring in Obs. 1 and there is not a significant dip in the UV flux in Obs. 3; it is not unusual to see variability in the X-rays, and none in the UV (e.g. see \citealt{Blu02} for analysis of \object{NGC 3783}).

\section{Preliminary modelling of the X-ray continuum and Fe K$\alpha$ line}
\label{continuum}

%
\begin{table*}
\begin{minipage}[t]{\hsize}
\setlength{\extrarowheight}{3pt}
\caption{Power-law and \FeKa emission line parameters, obtained from preliminary EPIC-pn fits (including the Galactic absorption) of all 4 observations over the 4.0--10 keV energy range.}
\label{pow_table}
\centering
\renewcommand{\footnoterule}{}
\begin{tabular}{c | c c | c c c c c c | c}
\hline \hline
 & \multicolumn{2}{c|}{Power-law:} & \multicolumn{6}{c|}{\FeKa:} & \\
{Obs.} & Photon Index ($\Gamma$) & Normalisation \footnote{$10^{51}$ photons $\mathrm{s}^{-1}$ $\mathrm{keV}^{-1}$ at 1 keV.} & $E_0$ \footnote{Theoretical rest frame wavelength in keV.} & $E$ \footnote{Wavelength in the rest frame of NGC 3516 in keV.} & Flow $v$ \footnote{\kms.} & FWHM \footnote {eV.} & $\sigma_{v}$ \footnote{RMS velocity in \kms ($\sigma_{v}=\mathrm{FWHM}/\sqrt{\ln 256}$).}& Normalisation \footnote{$10^{49}$ photons $\mathrm{s}^{-1}$.} & $\redchi$ / d.o.f. \\ 
\hline
1 & $1.85 \pm 0.02$ & $2.9 \pm 0.1$ & $6.39$ & $6.37 \pm 0.02$ & $+900_{-900}^{+1000}$ & $280 \pm 70$ & $6000 \pm 2000$ & $1.1 \pm 0.1$ & $1.54/330$\\
2 & $1.82 \pm 0.02$ & $2.5 \pm 0.1$ & $6.39$ & $6.38 \pm 0.01$ & $+500_{-500}^{+400}$ & $290 \pm 40$ & $6000 \pm 1000$ & $1.3 \pm 0.1$ & $1.42/361$\\
3 & $1.75 \pm 0.02$ & $1.9 \pm 0.1$ & $6.39$ & $6.38 \pm 0.01$ & $+500_{-500}^{+400}$ & $290 \pm 40$ & $6000 \pm 1000$ & $1.3 \pm 0.1$ & $1.58/324$\\
4 & $1.84 \pm 0.02$ & $2.6 \pm 0.1$ & $6.39$ &$ 6.40 \pm 0.01$ & $-500_{-400}^{+500}$ & $260 \pm 40$ & $5000 \pm 1000$ & $1.2 \pm 0.1$ & $1.38/311$\\
\hline
\end{tabular}
\end{minipage}
\end{table*}

\begin{figure}
\resizebox{\hsize}{!}{\includegraphics[angle=270]{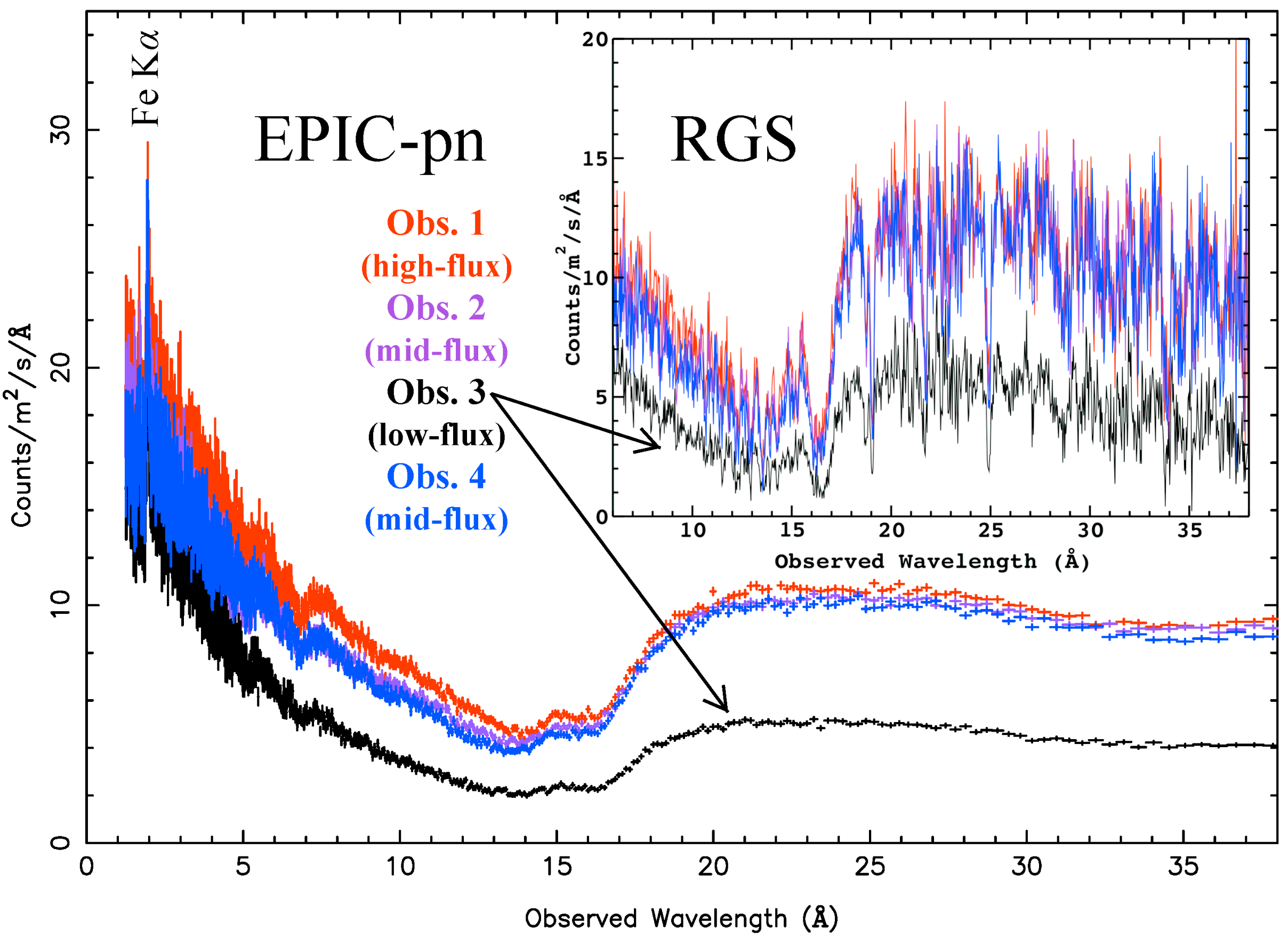}}
\caption{The EPIC-pn spectra of Obs. 1 (red), Obs. 2 (purple), Obs. 3 (black) and Obs. 4 (blue) binned to a minimum of 200 counts per bin. In the inset are the fluxed, combined RGS1 and RGS2 spectra binned by a factor of 3. Note that Obs. 2 and Obs. 4 are practically indistinguishable.}
\label{PN_all}
\end{figure}

Figure \ref{PN_all} depicts an overview of the EPIC-pn and RGS spectra of the 4 observations: while Obs. 2 and 4 display very similar spectra (so much so that they are practically indistinguishable), that of Obs. 1 lies slightly above them, and that of Obs. 3 is clearly much fainter than all the others. The features of the spectra, in particular during the dip in Obs. 3, are discussed in the following sections.

We started modelling the 4.0--10.0 keV EPIC-pn spectrum of Obs. 2 (which is the longest of the 4 observations) by a simple power-law ({\it pow}) redshifted to $z = 0.008836$. The transmission of the Galactic neutral absorption was included by applying the {\it hot} model (collisional ionisation equilibrium) in SPEX. The Galactic \ion{H}{i} column density in our line of sight was fixed to $N_{\mathrm{H}}={3.45\times 10^{20}\ \mathrm{cm}^{-2}}$ \citep{Kal05} and the gas temperature to 0.5 eV to mimic a neutral gas \citep{SPEX_man}. The prominent \FeKa emission line at $\sim$ 6.4 keV was modelled by a simple Gaussian profile ({\it gaus}) with its width left free to vary. The parameters obtained from the preliminary modelling of the continuum and Fe K$\alpha$ line are shown in Table \ref{pow_table}. Interestingly, we find that whereas the power-law flux falls significantly in Obs. 3, that of the \FeKa emission line remains unchanged compared to the other observations.

\begin{figure}
\resizebox{\hsize}{!}{\includegraphics[angle=90]{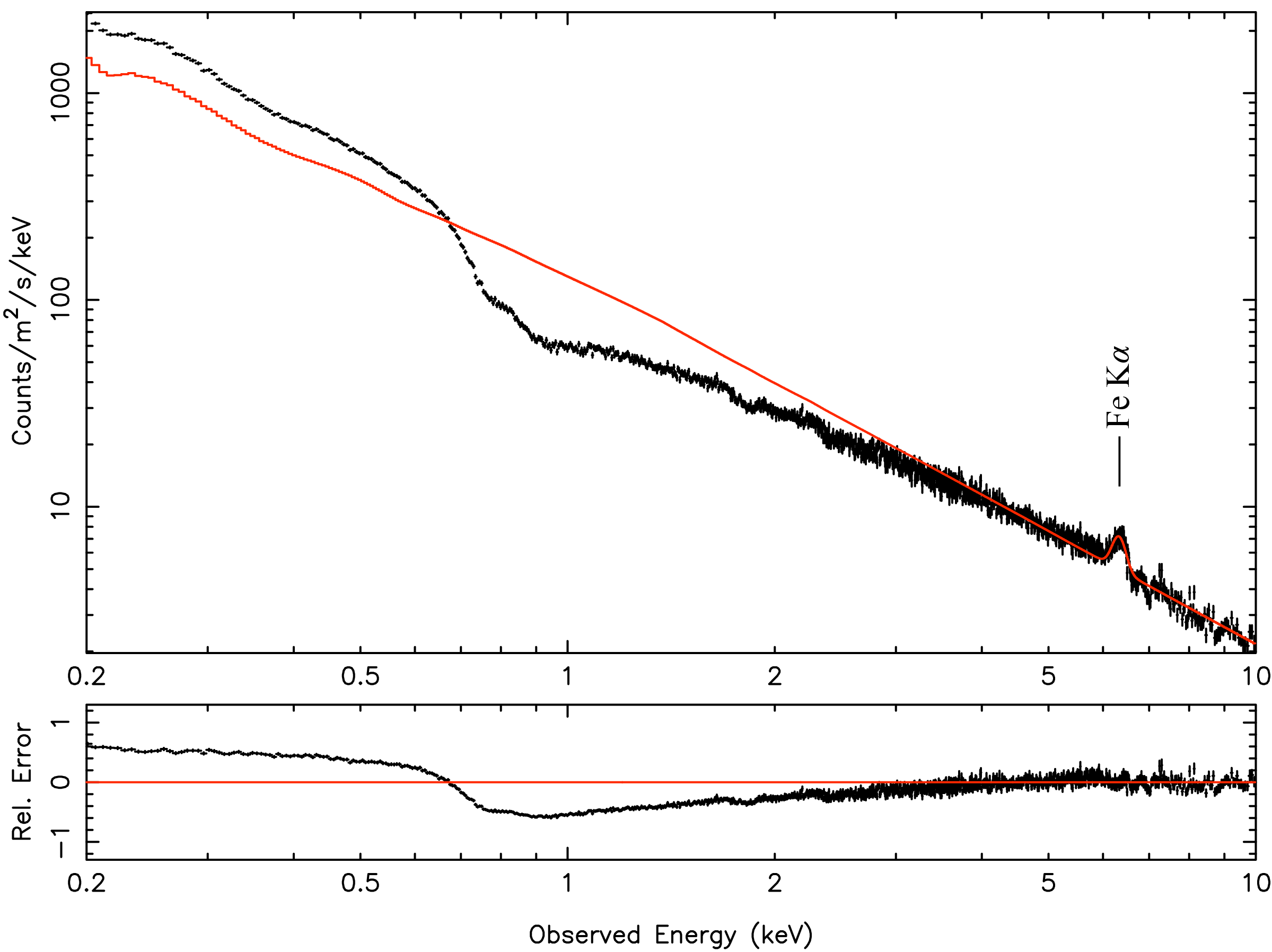}}
\caption{The Obs. 2 EPIC-pn power-law fit (including the Gaussian \FeKa line and Galactic absorption) over the 4.0--10.0 keV range. The fit is extrapolated to lower energies, displaying the presence of a soft excess and additional absorption in the 0.6--2.5 keV range.}
\label{PN_v1}
\end{figure}

Figure \ref{PN_v1} shows the 4.0--10.0 keV EPIC-pn fit, extrapolated to lower energies, displaying the presence of a soft excess and additional absorption in the 0.6--2.5 keV band. The latter is most likely due to the warm absorber and will be modelled in the next section, by fitting the RGS and EPIC-pn data simultaneously.

To include modelling of the soft excess, we refitted the EPIC-pn spectrum of Obs. 2 over the 0.2--10.0 keV energy range. We found that the power-law continuum model cannot fit the spectrum at all; a reduced Chi-squared (\redchi) value of 143 is obtained for 1111 degrees of freedom (d.o.f.). We then added a modified black body component ({\it mbb}) to the continuum in addition to the power-law. The {\it mbb} model describes the spectrum of a black body modified by coherent Compton scattering, and is often used to represent the accretion disk spectra of AGN \citep{Kaa89}. The addition of the {\it mbb} component ($kT=128\ \mathrm{eV}$) caused the \redchi to fall from 143 (1111 d.o.f.) to 21.0 (1109 d.o.f.). This is not yet a satisfactory fit, which shows the need for further modelling of the absorption in the 0.6--2.5 keV range (i.e. the warm absorber); this is discussed in the following section.

\section{Spectral modelling of the warm absorber}

In the following subsections we describe the process of fitting the spectra with a model including warm absorption. We start by fitting in detail the spectrum of Obs. 2, which represents an ``average'' state of the source. In Sect. \ref{slab_fit} we fit only the RGS spectrum of Obs. 2 and in Sect. \ref{xabs_fit} both the RGS and EPIC-pn simultaneously; we then proceed to fit the spectra of all the other observations. 

\subsection{Spectral fit using a thin slab absorption model}
\label{slab_fit}

\begin{figure}
\resizebox{\hsize}{!}{\includegraphics[angle=90]{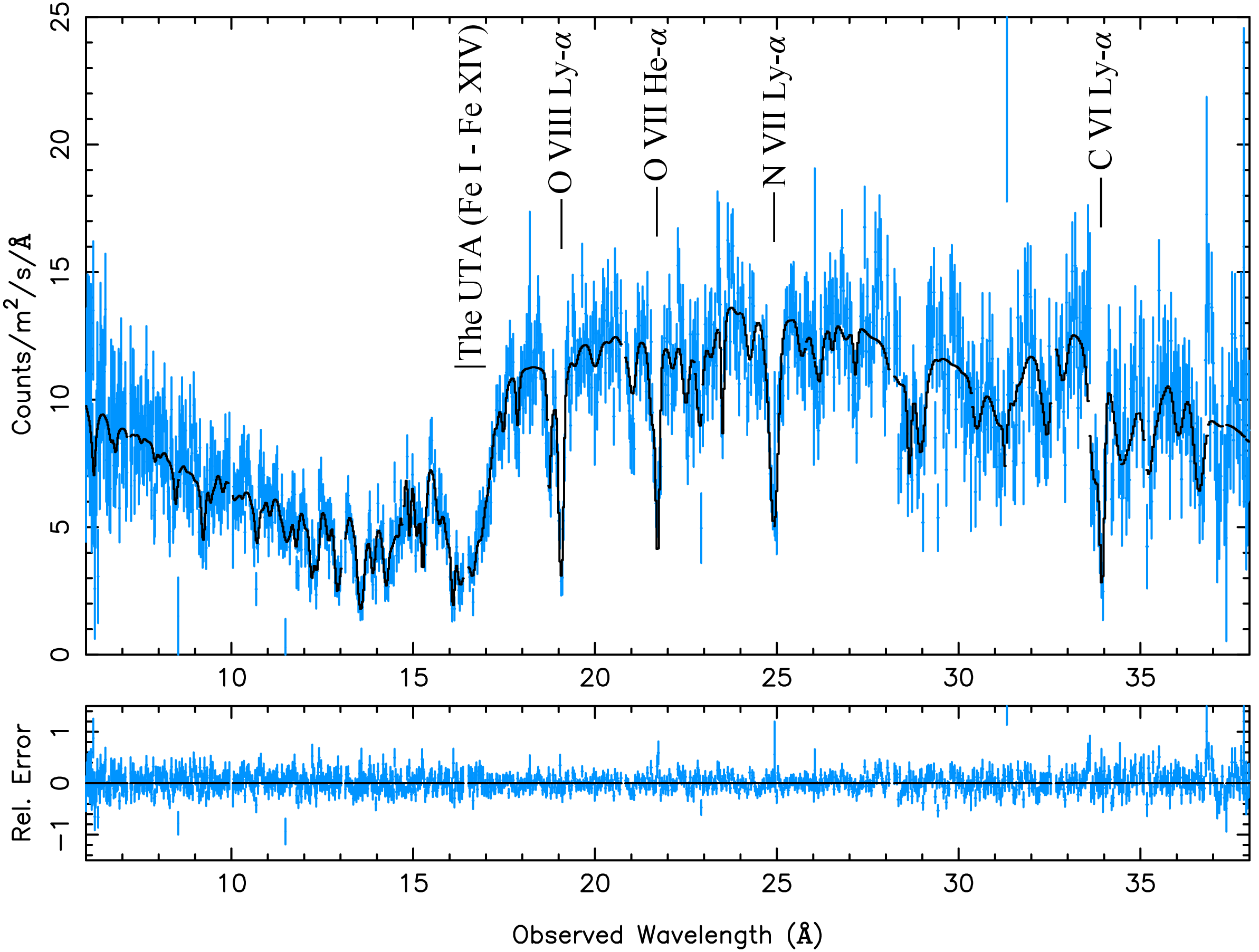}}
\caption{The RGS spectrum of Obs. 2, fitted using the \slab model described in Sect. \ref{slab_fit}. Some of the strongest absorption features are labelled.}
\label{rgs_whole}
\end{figure}

We began the analysis of the warm absorber by identifying and characterising the main absorption lines in the RGS spectrum of Obs. 2, shown in Fig. \ref{rgs_whole}. The {\it slab} model in SPEX calculates the transmission through a slab of material with adjustable ionic column densities. We started by adding a {\it slab} component to the best fit continuum model obtained from the EPIC-pn. All the continuum parameters except the power-law photon index were left free. The photon index was let free only after an overall good fit was obtained. This helps prevent being caught in false \redchi minima and obtaining a photon index incompatible with the EPIC-pn during the \redchi minimisation. This method of fixing the power-law photon index until close to the end of the fitting is also used in \citet{Ste03}.

%

\begin{table}
\setlength{\extrarowheight}{3pt}
\begin{minipage}[t]{\columnwidth}
\caption{Flow and RMS (broadening) velocities of the ions with the strongest absorption lines in the Obs. 2 RGS spectrum, determined from the {\it slab} model best fit.}
\label{slab_vel}
\centering
\renewcommand{\footnoterule}{}
\begin{tabular}{l l l l}
\hline \hline
{Ion} & Flow $v$ \footnote{km $\mathrm{s}^{-1}$.} & RMS $v$ $^a$ & Observed Transitions \\ 
\hline
\ion{C}{vi} & $-800 \pm 100$ & $300 \pm 100$ & Ly-$\alpha$, Ly-$\beta$, Ly-$\gamma$\\
\ion{N}{vi} & $-700 \pm 300$ & $600 \pm 200$ & He-$\alpha$, He-$\beta$, He-$\gamma$\\
\ion{N}{vii} & $-800 \pm 300$ & $900 \pm 300$ & Ly-$\alpha$, Ly-$\beta$, Ly-$\gamma$\\
\ion{O}{vii} & $-900 \pm 100$ & $300 \pm 100$ & He-$\alpha$, He-$\beta$, He-$\gamma$, He-$\delta$\\
\ion{O}{viii} & $-1000 \pm 100$ & $370 \pm 40$ & Ly-$\alpha$, Ly-$\beta$, Ly-$\gamma$, Ly-$\delta$\\
\hline
\end{tabular}
\end{minipage}
\end{table}

Initially all the ions in the \slab component have negligible column densities. We fitted the column density of each ion one by one, starting from the ions responsible for the strongest absorption lines in the spectrum. All the ions in a \slab component have the same flow and RMS (broadening) velocities. So after fitting the column densities of all the ions, we decoupled the velocities of some of the ions by creating a separate \slab component for each ion. Table \ref{slab_vel} shows the velocities determined for these ions. Only decoupling the velocities of the ions with strong absorption lines improves the fit. Decoupling velocities of the ions with weak lines does not improve the fit because large errors are associated with these ions. The ions with weak lines, that do not appear in Table \ref{slab_vel}, are modelled as one \slab component with an outflow velocity of $900 \pm 100$ \kms and RMS velocity of $900 \pm 100$ \kms.

We only used the \slab model to get some understanding of the scale of velocity and column density of the ions, especially those with the clearest absorption lines in the RGS spectrum (e.g. \ion{O}{viii}). To analyse the warm absorber in detail, we used a more realistic model described in the following subsection.   

\subsection{Spectral fit using a photoionised absorption model}
\label{xabs_fit}
The \xabs model in SPEX calculates the transmission through a slab of material, where all ionic column densities are linked in a physically consistent fashion through a grid of XSTAR \citep{Bau01} photoionisation models. We used the default \xabs model in which the Spectral Energy Distribution (SED) is based on the one from NGC 5548 as used in \cite{Ste05}. This is a good approximation for NGC 3516 since both Seyfert 1.5 AGN have very similar SEDs as shown in NASA/IPAC Extragalactic Database (NED). The advantage of the \xabs model over the \slab model is that all relevant ions are taken into account, including those, which would be detected only with marginal significance or not detected at all using the \slab model \citep{SPEX_man}.

We started with the EPIC-pn best fit continuum and \FeKa model established in Sect. \ref{continuum}, leaving the power-law, {\it mbb} and \FeKa parameters free (except the power-law photon index, which was kept fixed to the value given in Table \ref{pow_table} until close to the end of the fitting procedure). We applied the \xabs model to the continuum to fit the ionisation parameter ($\xi$), the hydrogen column density (\NH), the flow and RMS velocities in the EPIC-pn and RGS spectra of Obs. 2 simultaneously. The elemental abundances were fixed at the proto-solar values of \citet{Lod03}. Because we were fitting EPIC-pn, RGS1 and RGS2 spectra simultaneously, we allowed the relative normalisations of the EPIC-pn and the two RGS instruments to be free parameters. We fixed the instrumental normalisations of RGS1 to 1, and freed the normalisation of the RGS2 and EPIC-pn instruments. By doing this, problems with differences in relative normalisations between different instruments are avoided \citep{Blu03}. 

We then included one \xabs component in the model. This corresponds to phase B in Table \ref{xabs_table}, where the phases are sorted by increasing value of $\xi$. Adding the first \xabs component improved \redchi from 25.3 (2801 d.o.f.) to 3.91 (2797 d.o.f.), by fitting $\xi$, \NH, flow and RMS velocities. Our model still required more absorption at $\sim$11--15 $\AA$ (expected to be due to hot iron, \ion{Fe}{xvii}--\ion{Fe}{xxiv}) and $\sim$16--17 $\AA$ (most likely due to cold iron, \ion{Fe}{i}--\ion{Fe}{xiv}). So to produce a better fit to the data, we introduced a second \xabs component to our model. This corresponds to phase C in Table \ref{xabs_table}. The addition of phase C, which is the one with the highest ionisation parameter, further improved \redchi from 3.91 (2797 d.o.f.) to 1.88 (2793 d.o.f.). We found that after adding two \xabs components, absorption by iron, especially the M-shell iron (\ion{Fe}{i}--\ion{Fe}{xiv}) forming the Unresolved Transition Array (UTA) at $\sim$16--17 $\AA$ \citep{Beh01}, was still not properly modelled. To model the UTA, a phase with ionisation lower than the two previous ones identified was required. Therefore, a third \xabs component (phase A) was added to our model. The addition of phase A improved \redchi from 1.88 (2793 d.o.f.) to 1.63 (2789 d.o.f.) as the UTA was fitted well. At this point, the power-law photon index ($\Gamma$) was left free to vary in the fitting procedure for Obs. 2, but neither the value of $\Gamma$ (1.82) nor \redchi changed significantly.

To attempt to improve further the quality of our fit, we freed the abundances of only the elements with the strongest absorption features in the RGS spectrum: N, O and Fe. In order not to increase the number of free parameters unnecessarily, we coupled the abundances of all three phases. The precise abundances of elements in each phase cannot be accurately determined since letting them all free results in some cases in unphysical values. Assuming that different phases have the same abundances is already a more progressive approach than the alternative of assuming cosmic abundances for all the phases. The abundance coupling approach is also used by \citet{Kaa03_2} where they fixed the abundances of a cool gas component to those of a hot one, in order to reduce the number of free parameters. Freeing the N, O and Fe abundances, produced a better fit as \redchi fell from 1.63 (2788 d.o.f.) to 1.52 (2785 d.o.f.).

\begin{figure}[!]
\resizebox{\hsize}{!}{\includegraphics[angle=270]{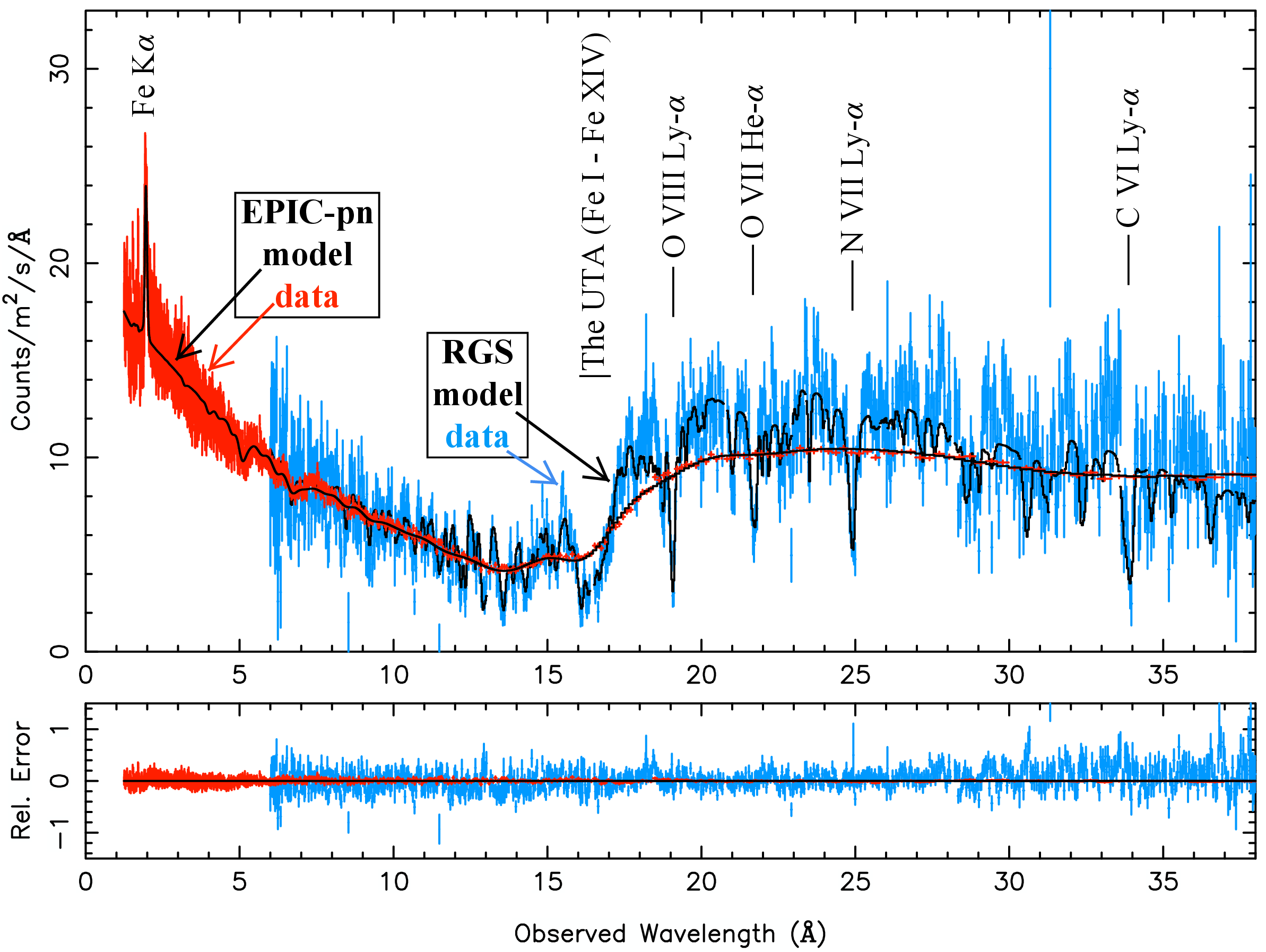}}
\caption{The Obs. 2 best fit \xabs model, fitted simultaneously to the RGS and EPIC-pn spectra. The RGS data are shown in blue, the EPIC-pn data in red and the models in black. Close-ups of the RGS spectrum, showing the spectral features more clearly, are presented in Fig. \ref{ovii_caxiv}. Note that both RGS and EPIC-pn spectra in this figure have been divided by the instrument's effective area, thus are plotted on the same scale.}
\label{RGS_PN_whole}
\end{figure}

One of the ionisation phases in \cite{Tur08} with $\log \xi$ of 2.2 has a partial covering fraction. Therefore, at this stage of the fitting process we tested whether any of our three phases is partially covering the source. We freed the covering fraction (\fcov) of the three phases from the default value of 1. The covering fraction of one of the phases (Phase B with $\log \xi$ of 2.4) was found to change from 1 to 0.55, whist the \fcov of the other two phases remained practically unchanged. By freeing the covering fraction of phase B, \redchi fell from 1.52 (2785 d.o.f.) to 1.44 (2784 d.o.f.). In fact there are absorption lines in the 11--13 $\AA$ region of the RGS spectrum which are not fitted well if \fcov is fixed to 1, as well as the UTA and the \ion{O}{vii} He-$\alpha$ line. So a partial covering of the continuum is necessary for phase B, but this is not the case for the other two phases. Therefore, in the following we allowed phase B to have a partial covering fraction and \fcov of phases A and C were fixed to 1. A plot of the whole RGS spectrum, the EPIC-pn data and the best fit model for Obs. 2 is shown in Fig. \ref{RGS_PN_whole}.
%
\begin{landscape}
\begin{table}
\caption{Best fit parameters of the continuum,
obtained from the three-phase {\it xabs} model simultaneous fit to the EPIC-pn, RGS1 and RGS2 spectra.}
\label{cont_table}
\setlength{\extrarowheight}{3pt}
\begin{minipage}[t]{\columnwidth}
\renewcommand{\footnoterule}{}
\centering
\begin{tabular}{c | l l | l l}
\hline \hline
 & \multicolumn{2}{c|}{Power-law:} & \multicolumn{2}{c}{Modified black body:} \\ 
Obs. & Photon Index ($\Gamma$) & Normalisation \footnote{$10^{51}$ photons $\mathrm{s}^{-1}$ $\mathrm{keV}^{-1}$ at 1 keV.} & Temperature \footnote{eV.} & Normalisation \footnote{Emitting area times square root of the electron density in $10^{33}$ $\mathrm{cm}^{0.5}$.} \\
\hline

1 & $1.85 \pm 0.02$ & $3.3 \pm 0.1$ & $189 \pm 2$ & $1.3 \pm 0.1$ \\

2 & $1.82 \pm 0.01$ & $2.8 \pm 0.1$ & $186 \pm 3$ & $1.5 \pm 0.1$ \\

3 & $1.70 \pm 0.02$ & $1.9 \pm 0.1$ & $211 \pm 4$ & $0.5 \pm 0.1$ \\

4 & $1.84 \pm 0.02$ & $2.9 \pm 0.1$ & $189 \pm 2$ & $1.4 \pm 0.1$ \\
\hline
\end{tabular}
\end{minipage}
\end{table}
\begin{table}
\setlength{\extrarowheight}{3pt}
\begin{minipage}[t]{\columnwidth}
\caption{Best fit parameters of the three-phase {\it xabs} model for all four observations, obtained from the simultaneous fitting of the EPIC-pn, RGS1 and RGS2 spectra.}
\label{xabs_table}
\tiny
\centering
\renewcommand{\footnoterule}{}
\begin{tabular}{c | l l l l | l l l l l | l l l l | l}
\hline \hline
 & \multicolumn{4}{c|}{Phase A:} & \multicolumn{5}{c|}{Phase B:} & \multicolumn{4}{c|}{Phase C:} & \\ 

Obs. & log $\xi$ \footnote{$\mathrm{erg\ cm\ }\mathrm{s}^{-1}$.} & $N_{\mathrm{H}}$ \footnote{$10^{22}$ $\mathrm{cm}^{-2}$.} & Flow $v$ \footnote{km $\mathrm{s}^{-1}$.} & RMS $v$ $^c$

& log $\xi$ $^a$ & $N_{\mathrm{H}}$ $^b$ & Flow $v$ $^c$ & RMS $v$ $^c$ & {\it fcov} \footnote{The covering fraction of Phases A and C is 1.0.} 

& log $\xi$ $^a$ & $N_{\mathrm{H}}$ $^b$ & Flow $v$ $^c$ & RMS $v$ $^c$

& $\chi _\nu ^2$ / d.o.f. \footnote{The \redchi values are further improved by including the modelling of the observed emission features discussed in Sect. \ref{emission_sec}.} \\

\hline

1 & $0.95 \pm 0.02$ & $0.33 \pm 0.01$ & $-100 \pm 40$ & $60 \pm 10$ &

 $2.41 \pm 0.05$ & $1.7 \pm 0.2$ & $-1500 \pm 100$ & $500 \pm 100$ & $0.64 \pm 0.09$ & 
 
 $3.00 \pm 0.05$ & $1.2 \pm 0.2$ & $-800 \pm 300$ & $300 \pm 100$ & $1.41/2746$ \\

2 & $0.96 \pm 0.02$ & $0.34 \pm 0.01$ & $-100 \pm 30$ & $40 \pm 10$ &

 $2.42 \pm 0.04$ & $2.0 \pm 0.3$ & $-1500 \pm 100$ & $500 \pm 100$ & $0.55 \pm 0.04$ & 
 
 $2.99 \pm 0.03$ & $1.3 \pm 0.2$ & $-1000 \pm 100$ & $400 \pm 100$ & $1.44/2784$ \\

3 & $0.87 \pm 0.02$ & $0.43 \pm 0.02$ & $-200 \pm 40$ & $40 \pm 10$ &

 $2.43 \pm 0.03$ & $3.2 \pm 0.4$ & $-1600 \pm 200$ & $400 \pm 100$ & $0.64 \pm 0.04$ & 
 
 $3.07 \pm 0.04$ & $1.7 \pm 0.2$ & $-800 \pm 200$ & $400 \pm 100$ & $1.41/2689$ \\

4 & $0.97 \pm 0.02$ & $0.37 \pm 0.01$ & $-200 \pm 20$  & $30 \pm 10$ &

 $2.39 \pm 0.05$ & $2.0 \pm 0.1$ & $-1500 \pm 100$ & $600 \pm 100$ & $0.54 \pm 0.04$ & 
 
 $2.99 \pm 0.02$ & $1.7 \pm 0.2$ & $-1000 \pm 100$ & $500 \pm 100$ & $1.42/2724$ \\[2pt]

\hline
\end{tabular}
\end{minipage}
\end{table}
\begin{table}
\setlength{\extrarowheight}{3pt}
\begin{minipage}[t]{\columnwidth}
\caption{The elemental abundances of the three-phase {\it xabs} model of Obs. 2 fitted simultaneously to the EPIC-pn, RGS1 and RGS2 spectra. The abundances of the three phases were coupled to each other in the fits. All the abundances are relative to the proto-solar model of \citet{Lod03}. The reference element is H.}
\label{abun_table}
\centering
\renewcommand{\footnoterule}{}
\begin{tabular}{c c}
\hline \hline
{Element} & {Abundance relative to H} \\ 
\hline
N & $1.5 \pm 0.1$\\
O & $1.4 \pm 0.1$\\
Fe & $1.8 \pm 0.1$\\
Other elements & 1 (f)\\[2pt]
\hline
\end{tabular}
\end{minipage}
\end{table}
\end{landscape}
\begin{figure}
\resizebox{\hsize}{!}{\includegraphics[angle=270]{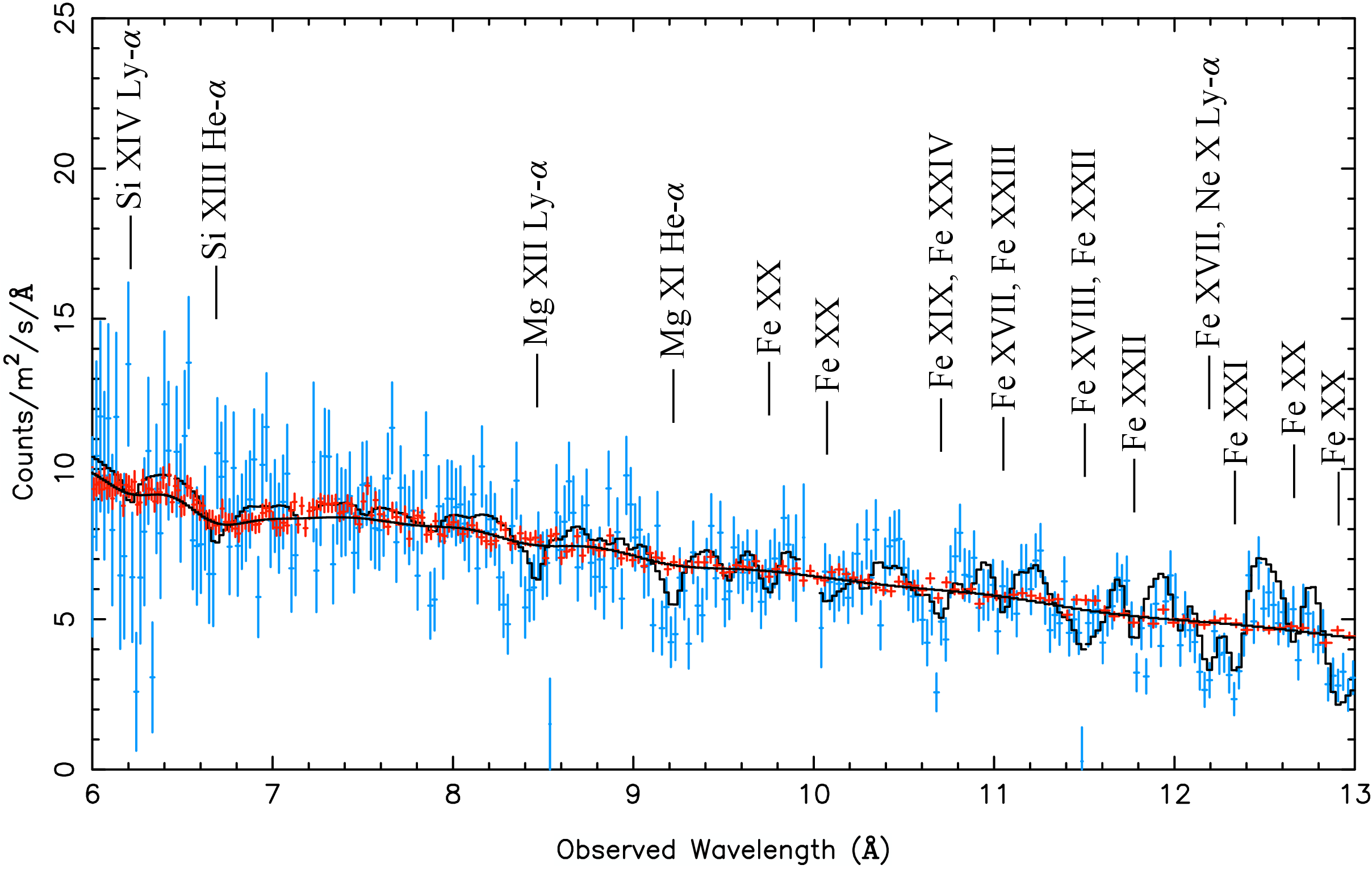}}
\resizebox{\hsize}{!}{\includegraphics[angle=270]{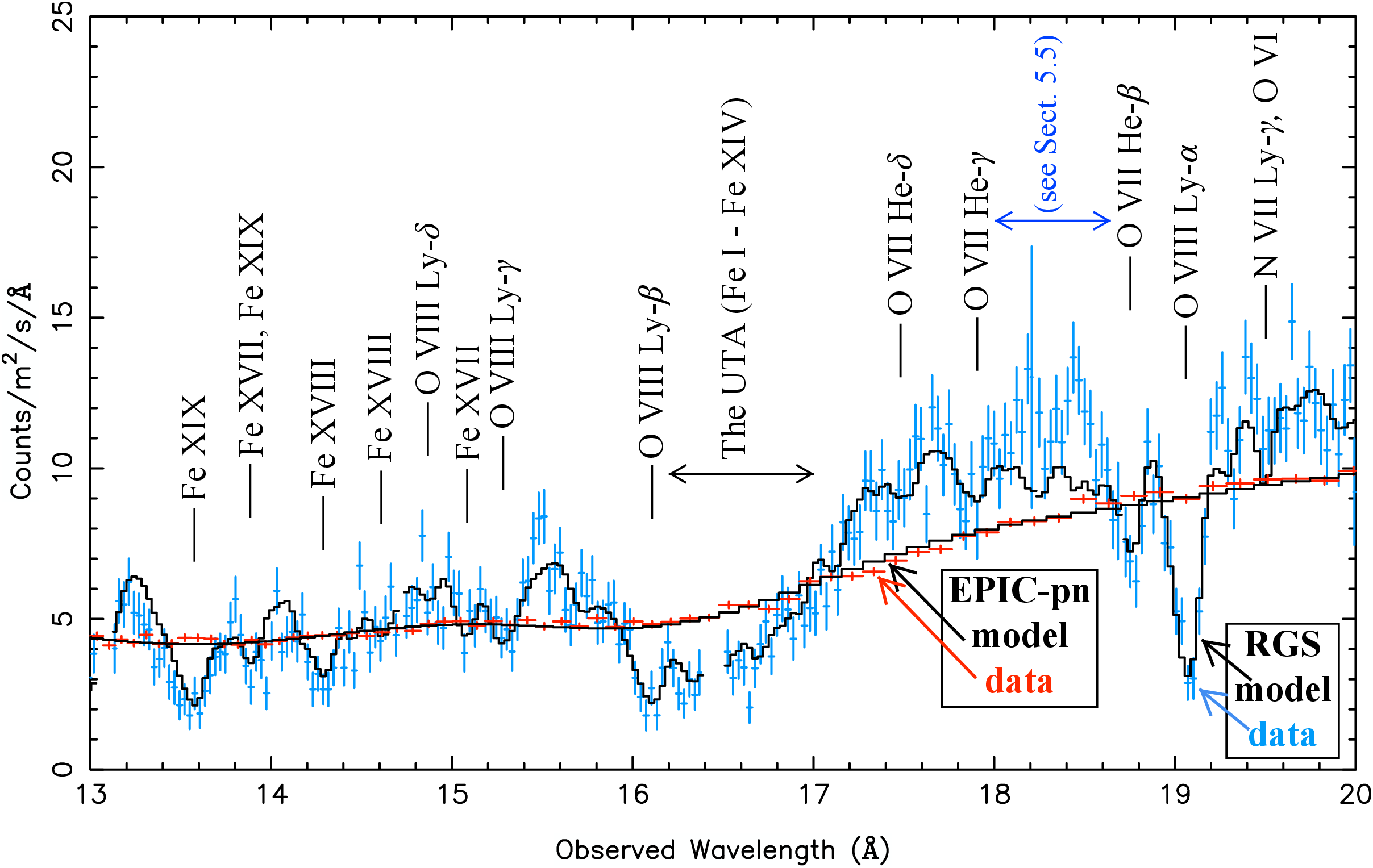}}
\resizebox{\hsize}{!}{\includegraphics[angle=270]{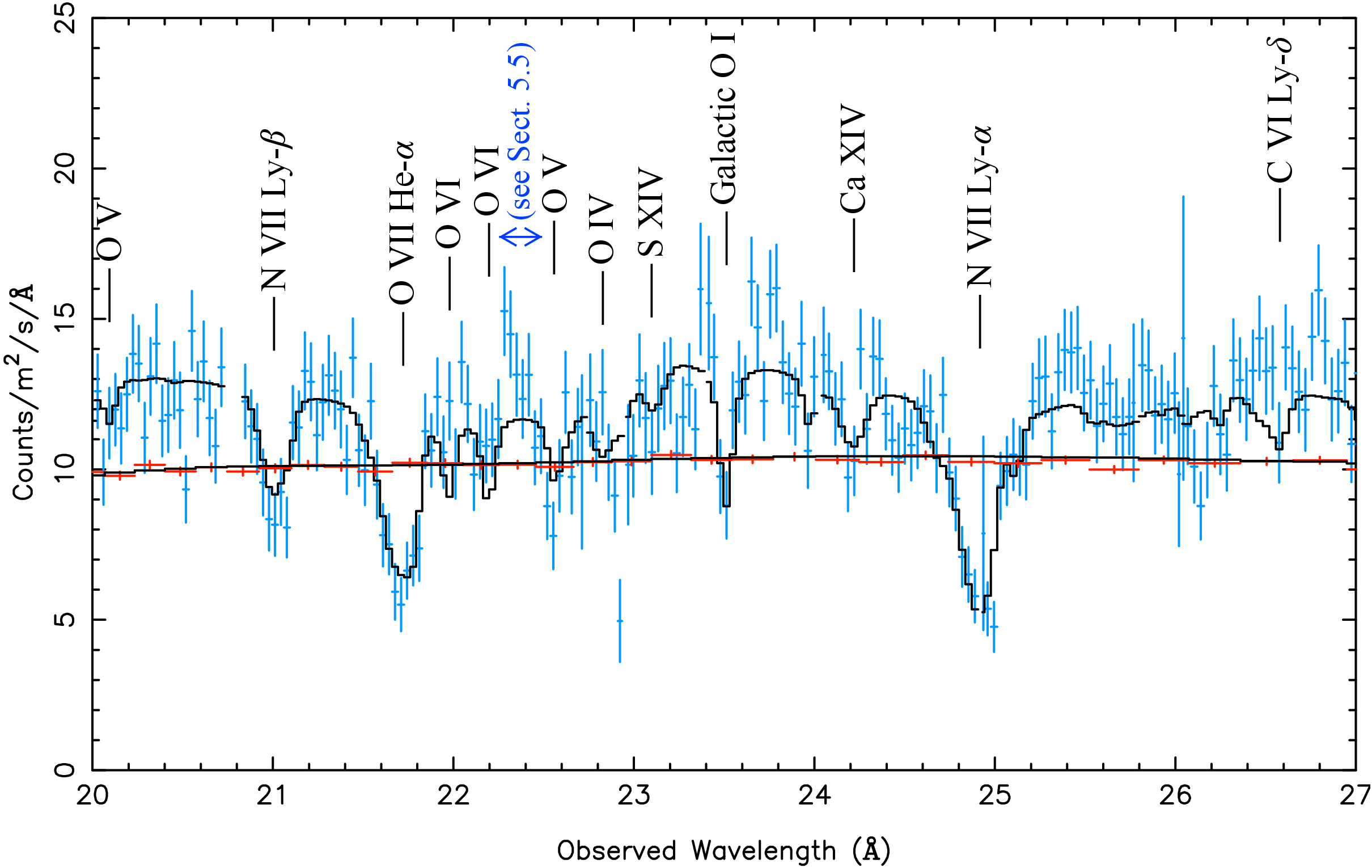}}
\resizebox{\hsize}{!}{\includegraphics[angle=270]{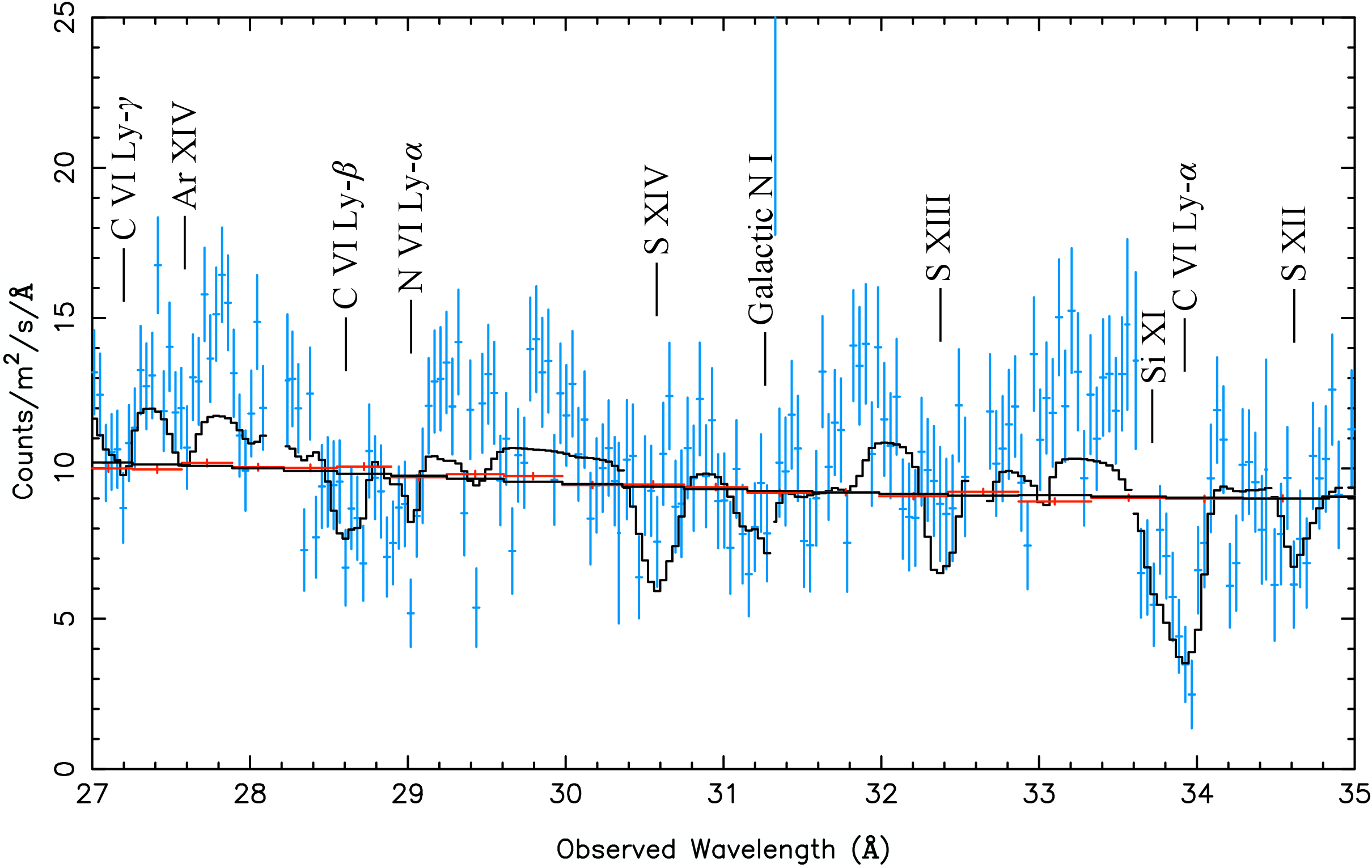}}
\caption{The Obs. 2 best fit \xabs model, fitted simultaneously to the RGS and EPIC-pn spectra. The RGS data are shown in blue, the EPIC-pn data in red and the models in black.}
\label{ovii_caxiv}
\end{figure}

\newpage

The best fit parameters for Obs. 2 (power-law, {\it mbb}, \xabs and abundances) are listed in Tables \ref{cont_table}, \ref{xabs_table} and \ref{abun_table}. In Fig. \ref{ovii_caxiv}, close-ups of the RGS spectrum and the best fit model are presented. Having obtained the \xabs model best fit for Obs. 2, we used it as the starting point for each of the other observations, leaving all the parameters quoted so far free, with the exception of the abundances of the \xabs phases which were fixed at the Obs. 2 best fit (the abundance values are shown in Table \ref{abun_table}). The best fit parameters of the continuum and ionisation phases of Obs. 1, Obs. 3 and Obs. 4, modelled with the \xabs photoionised absorption, are also shown in Tables \ref{cont_table} and \ref{xabs_table}. Note that the \redchi values in Table \ref{xabs_table} are further improved by modelling the observed emission features discussed in Sect. \ref{emission_sec}. In the following subsection (Sect. \ref{obs3}), we describe how the Obs. 2 best fit was applied to the Obs. 3 spectrum, in order to investigate whether the covering fraction of phase B is variable or not.

\subsection{Spectral variability in Obs. 3}
\label{obs3}

As shown in Table \ref{xabs_table}, phase B in Obs. 2 has a best fit covering fraction (\fcov) of 0.55. Our phase B is very similar to the partially covering phase (zone 3) of \citet{Tur08} in terms of ionisation parameter. We find $\log\ \xi \sim 2.4$ versus \citeauthor{Tur08}'s $\log\ \xi \sim 2.2$. However, the column density of our phase B ($\NH \sim 2.0 \times 10^{22}\ \mathrm{cm}^{-2}$) is about a factor of 10 smaller than the one in \citet{Tur08} ($\NH \sim 20 \times 10^{22}\ \mathrm{cm}^{-2}$). In Sect. \ref{abs_discu}, we discuss the values of \NH of warm absorber phases in NGC 3516 found by different authors.

\begin{figure}[!]
\resizebox{\hsize}{!}{\includegraphics[angle=270]{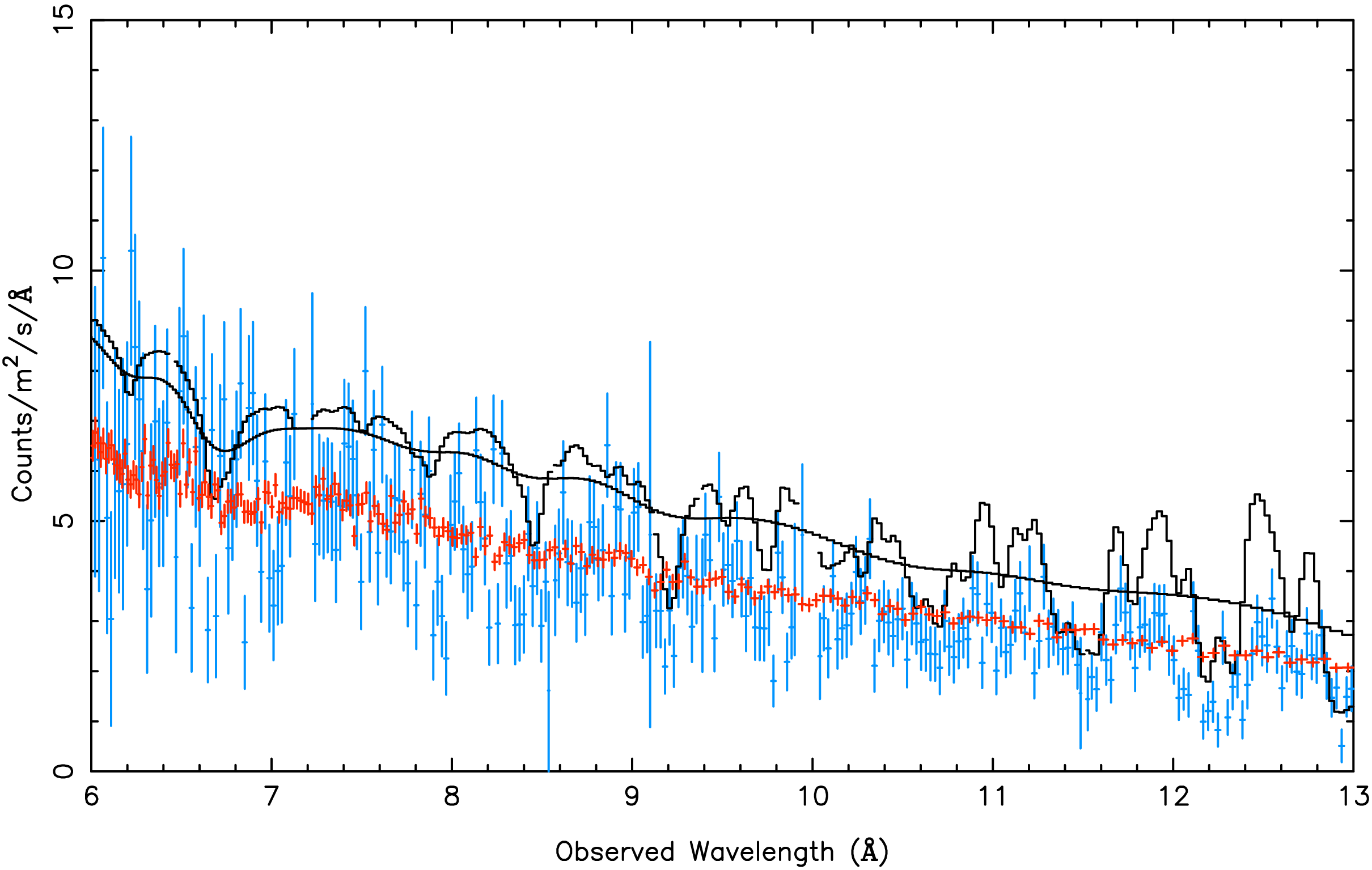}}
\resizebox{\hsize}{!}{\includegraphics[angle=270]{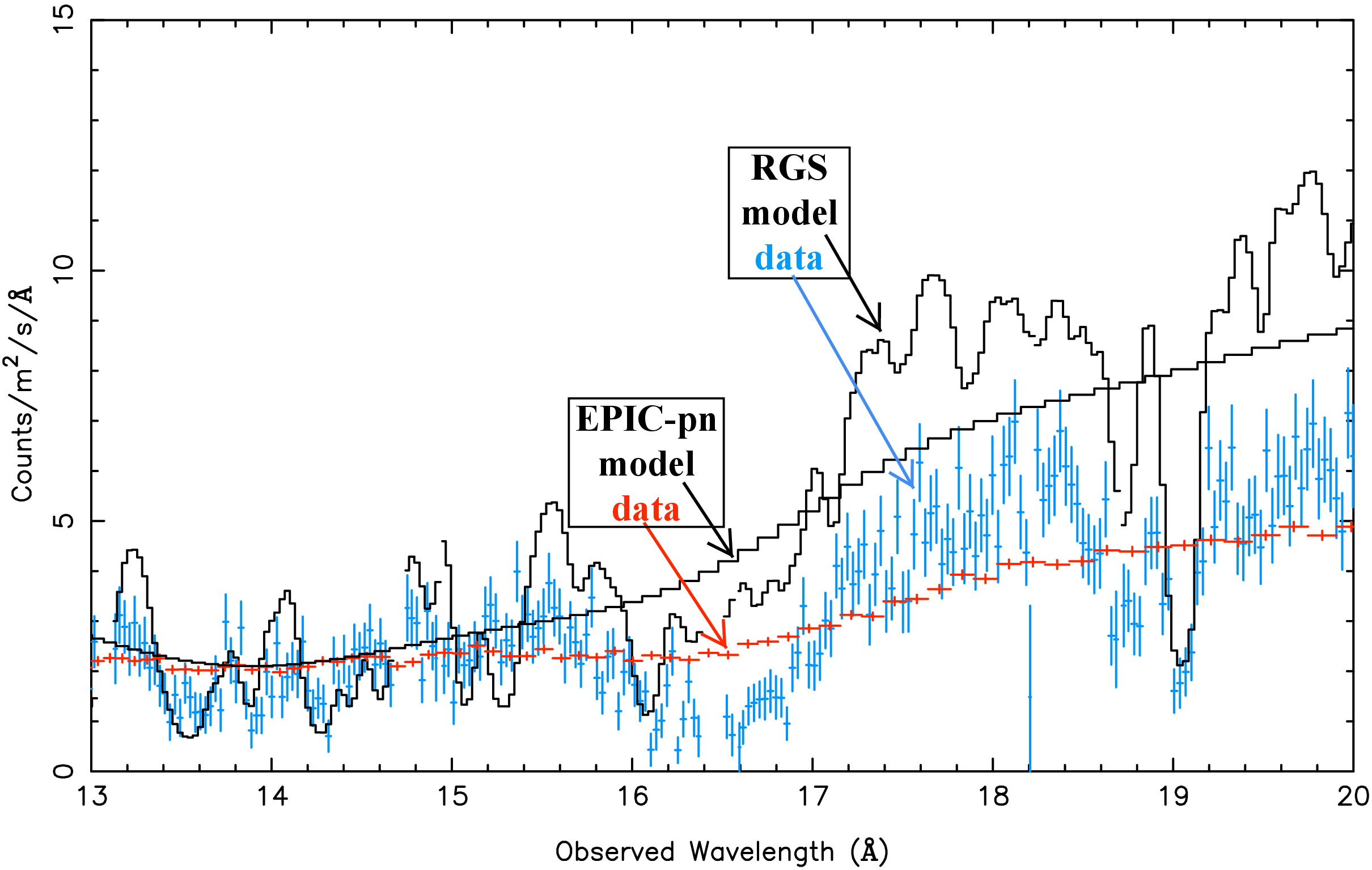}}
\resizebox{\hsize}{!}{\includegraphics[angle=270]{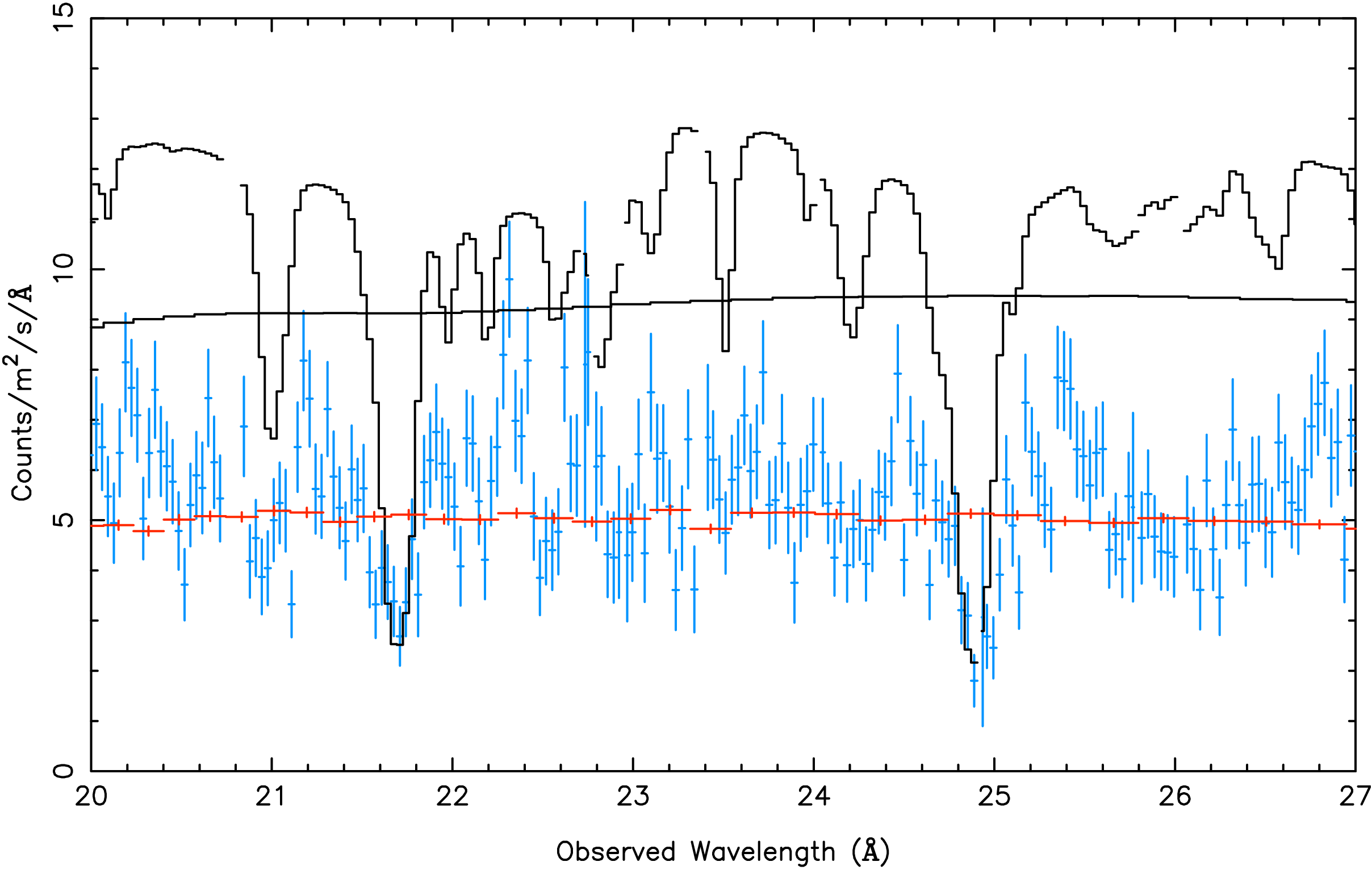}}
\resizebox{\hsize}{!}{\includegraphics[angle=270]{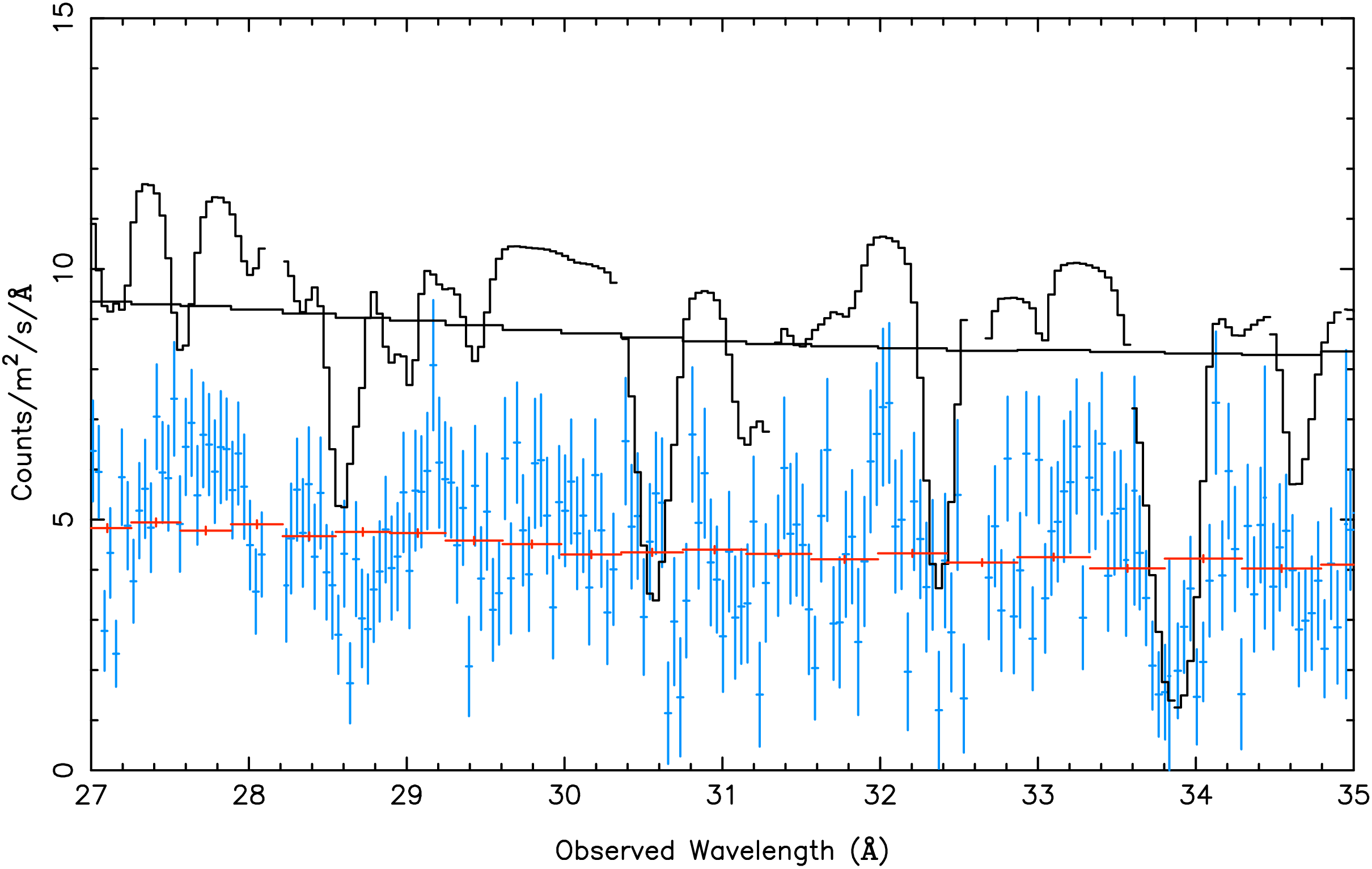}}
\caption{The simultaneous fit made to the EPIC-pn and RGS spectra of Obs. 3, by using the best fit model of Obs. 2 and fitting only the covering fraction of phase B. The \redchi is 130 (2710 d.o.f.). The RGS data are shown in blue, the EPIC-pn data in red and the models in black.}
\label{bad_fit}
\end{figure}

\begin{figure}[!]
\resizebox{\hsize}{!}{\includegraphics[angle=270]{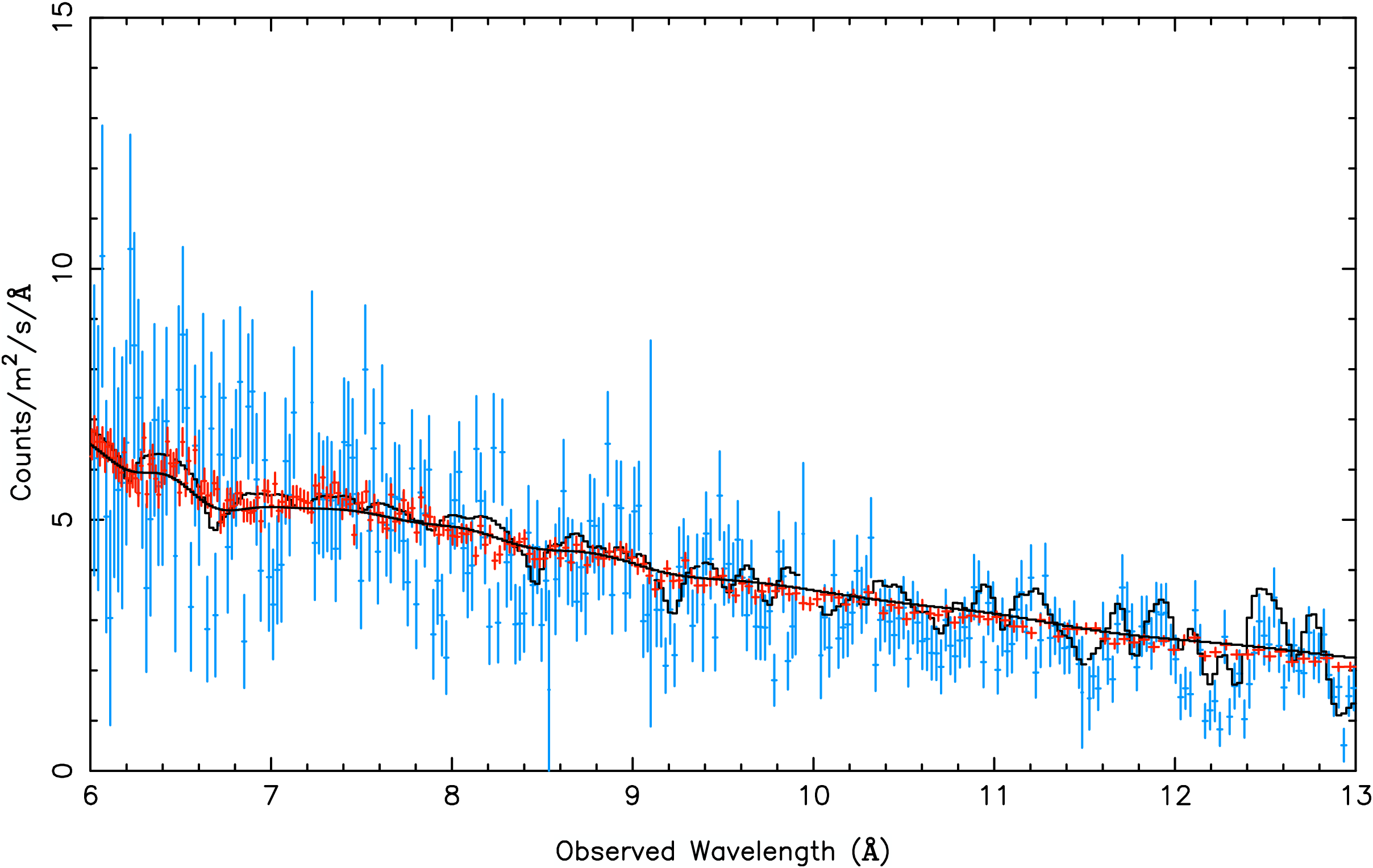}}
\resizebox{\hsize}{!}{\includegraphics[angle=270]{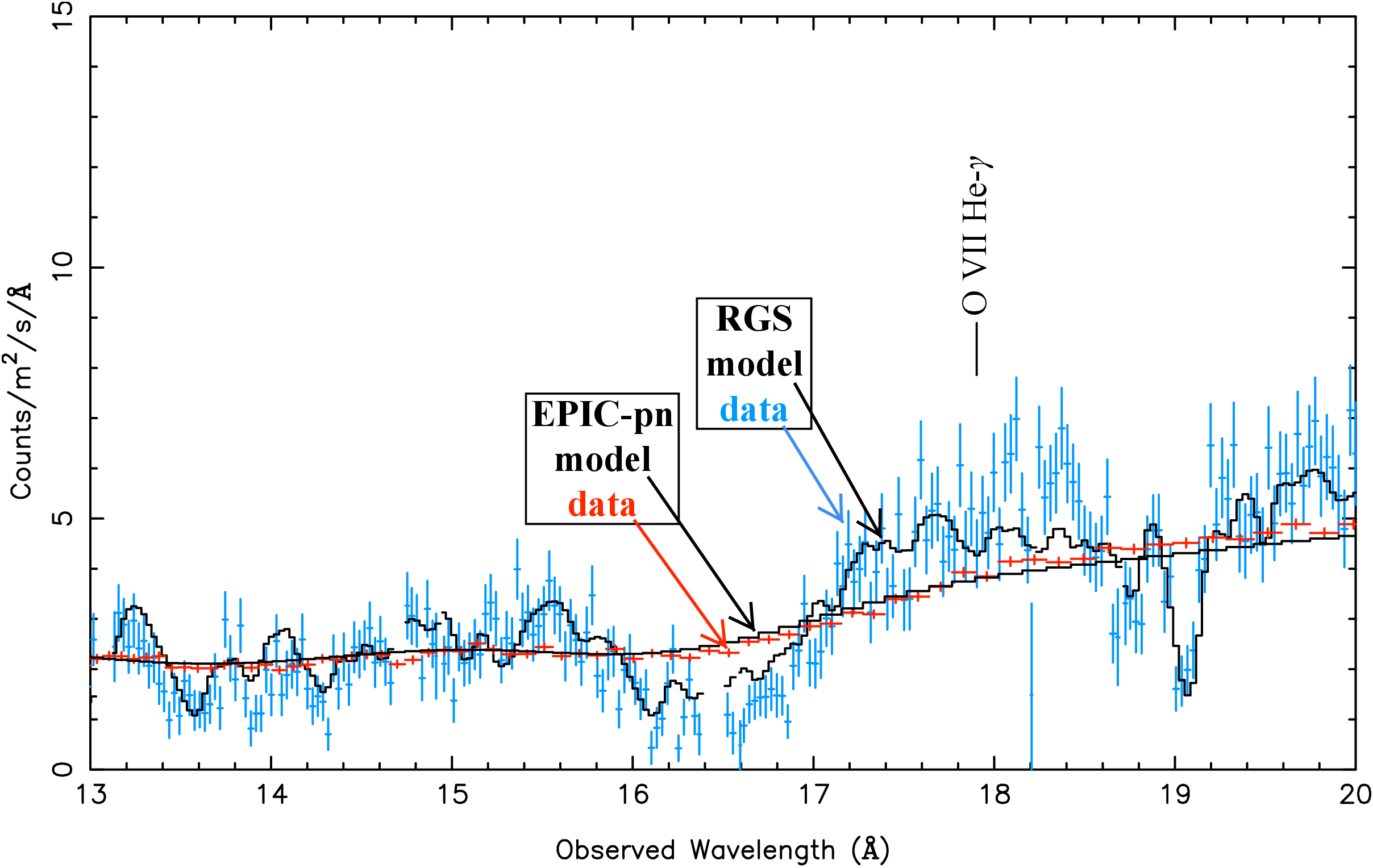}}
\resizebox{\hsize}{!}{\includegraphics[angle=270]{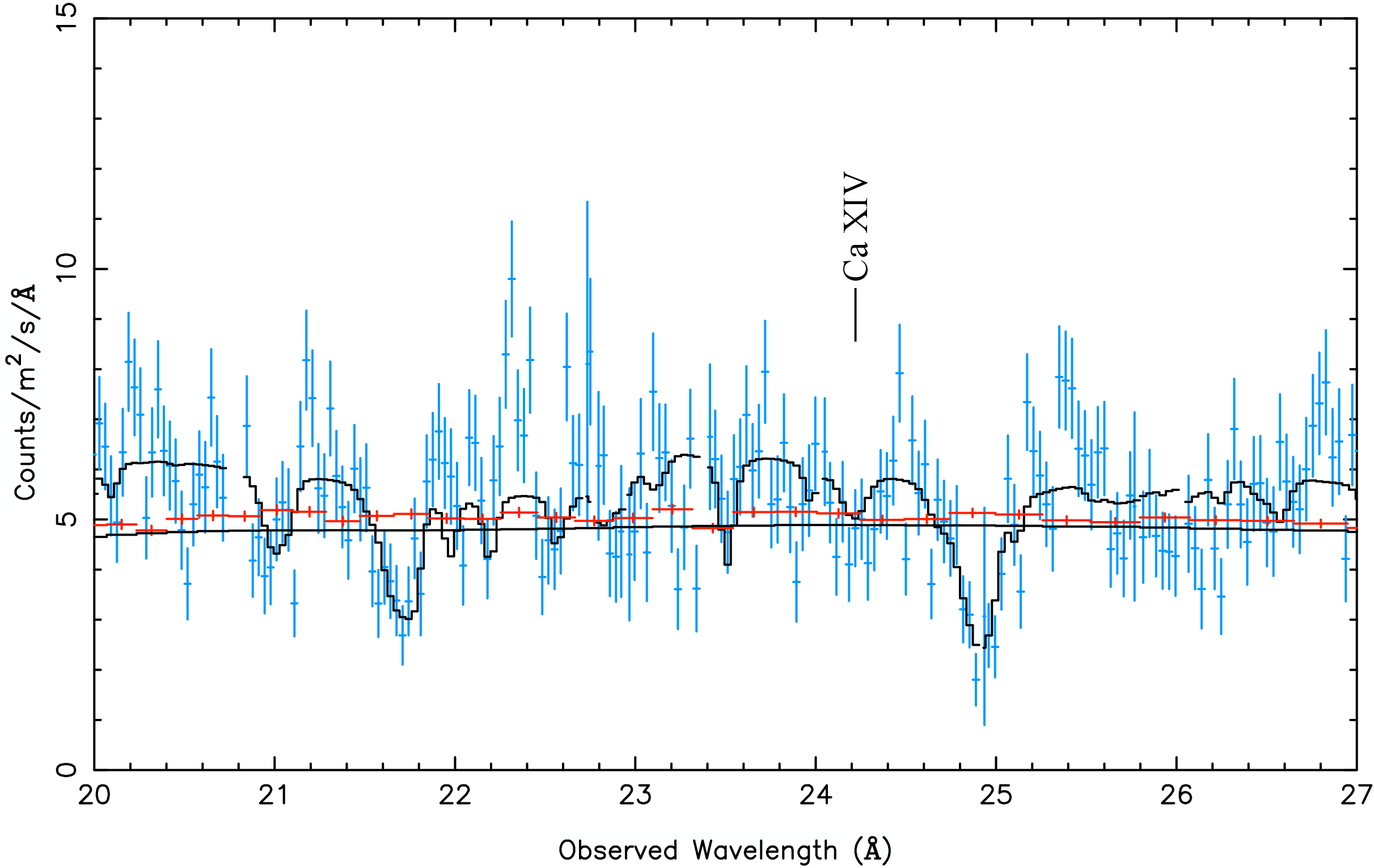}}
\resizebox{\hsize}{!}{\includegraphics[angle=270]{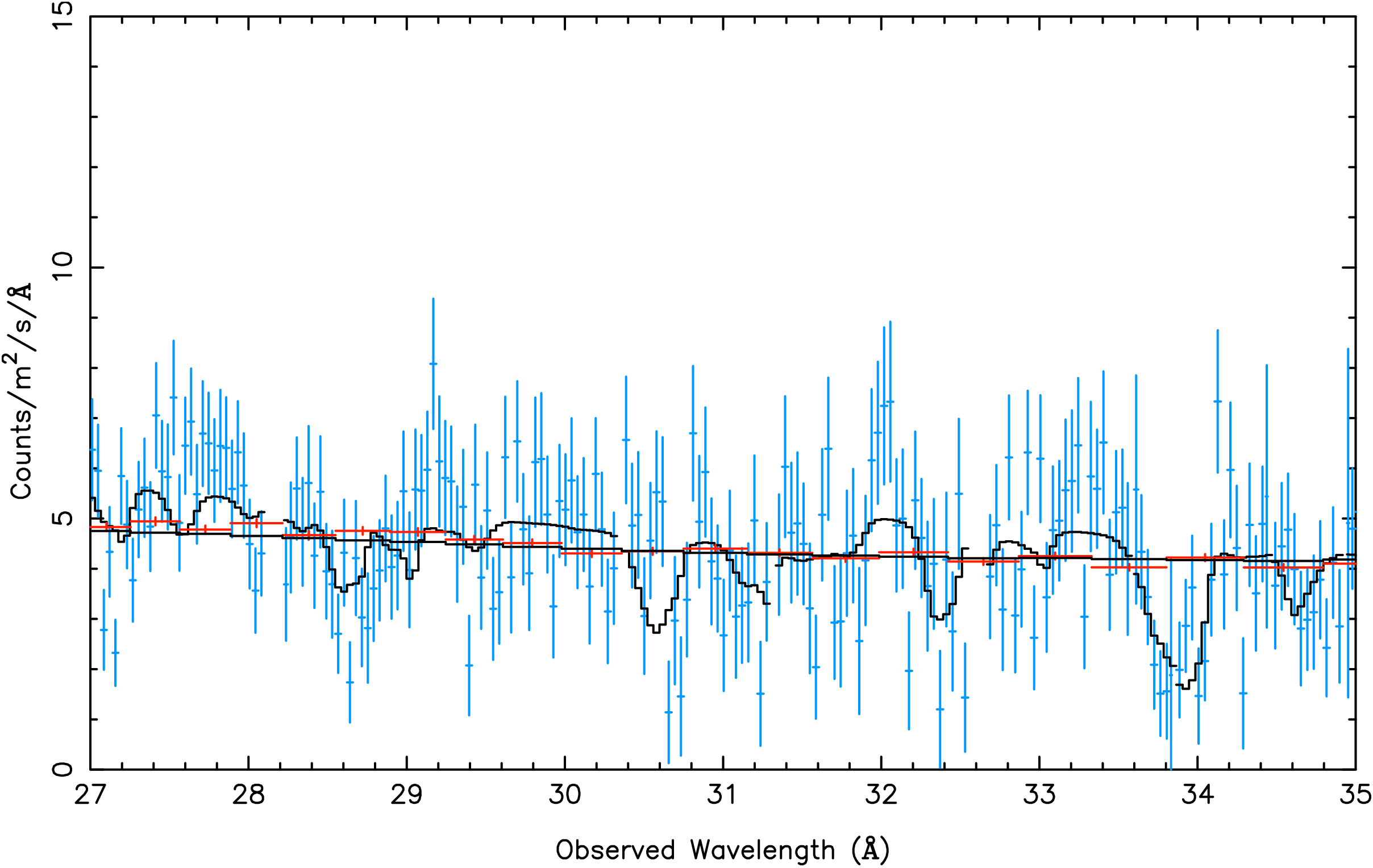}}
\caption{The simultaneous fit made to the EPIC-pn and RGS spectra of Obs. 3, using the best fit model of Obs. 2 and fitting only the continuum parameters (maintaining $\fcov=0.55$). The \redchi is 1.76 (2707 d.o.f.). The RGS data are shown in blue, the EPIC-pn data in red and the models in black. The significance of the \ion{O}{vii} He-$\gamma$ and \ion{Ca}{xiv} lines is discussed in Sect. \ref{three_phase}.}
\label{good_fit}
\end{figure}

\citet{Tur08} explained the dip in the X-ray lightcurve of Obs. 3 by varying only the covering fraction of their zone 3. In order to try and reproduce their results, we applied our Obs. 2 best fit (described in Sect. \ref{xabs_fit}) to Obs. 3 by keeping all the parameters fixed and freeing only the covering fraction. As for Obs. 2, we fitted the EPIC-pn and RGS spectra simultaneously. Applying the Obs. 2 best fit to Obs. 3 we obtained an initial \redchi of 225 (2711 d.o.f.). We found that by freeing only the covering fraction no good fit was achieved; the covering fraction goes to 1 and a \redchi of 130 (2710 d.o.f.) is obtained. On the other hand, when we freed only the continuum parameters, a much better fit was obtained, \redchi of 1.76 (2707 d.o.f.). Figures \ref{bad_fit} and \ref{good_fit} depict the difference between the two approaches to Obs. 3, proving that changing the continuum parameters is a much more feasible solution than changing only the covering fraction of phase B in order to fit the Obs. 3 spectrum.


We then tried freeing both the continuum and covering fraction. In this case, the covering fraction becomes 0.70 and \redchi goes to 1.65 (2706 d.o.f.). We then freed the ionisation parameter and \NH of the three \xabs phases, parameters of the \FeKa line and the relative normalisations of the EPIC-pn and RGS instruments; the \redchi improved to 1.42 (2695 d.o.f.). Finally, we freed the velocities of the \xabs phases, keeping the abundances fixed. In this case, the fit does not improve significantly, a \redchi of 1.41 (2689 d.o.f.) was obtained. The best fit results for Obs. 3, as for the other three observations, are shown in Tables \ref{cont_table} and \ref{xabs_table}.

It is worth noting the advantage of modelling both the RGS and EPIC-pn together, which leads to simultaneously constraining the broadband continuum using the EPIC-pn and modelling the absorption features using the RGS. This approach has a drawback in that, due to smaller statistical fractional errors associated with the EPIC-pn compared to the RGS, the best fit favours the EPIC-pn. However, by making sure that the RGS absorption features are fitted properly, this weakness is not a problem.

\subsection{The partially covering phase B}
\label{three_phase}

\begin{figure}
\centering
\resizebox{\hsize}{!}{\includegraphics[angle=270]{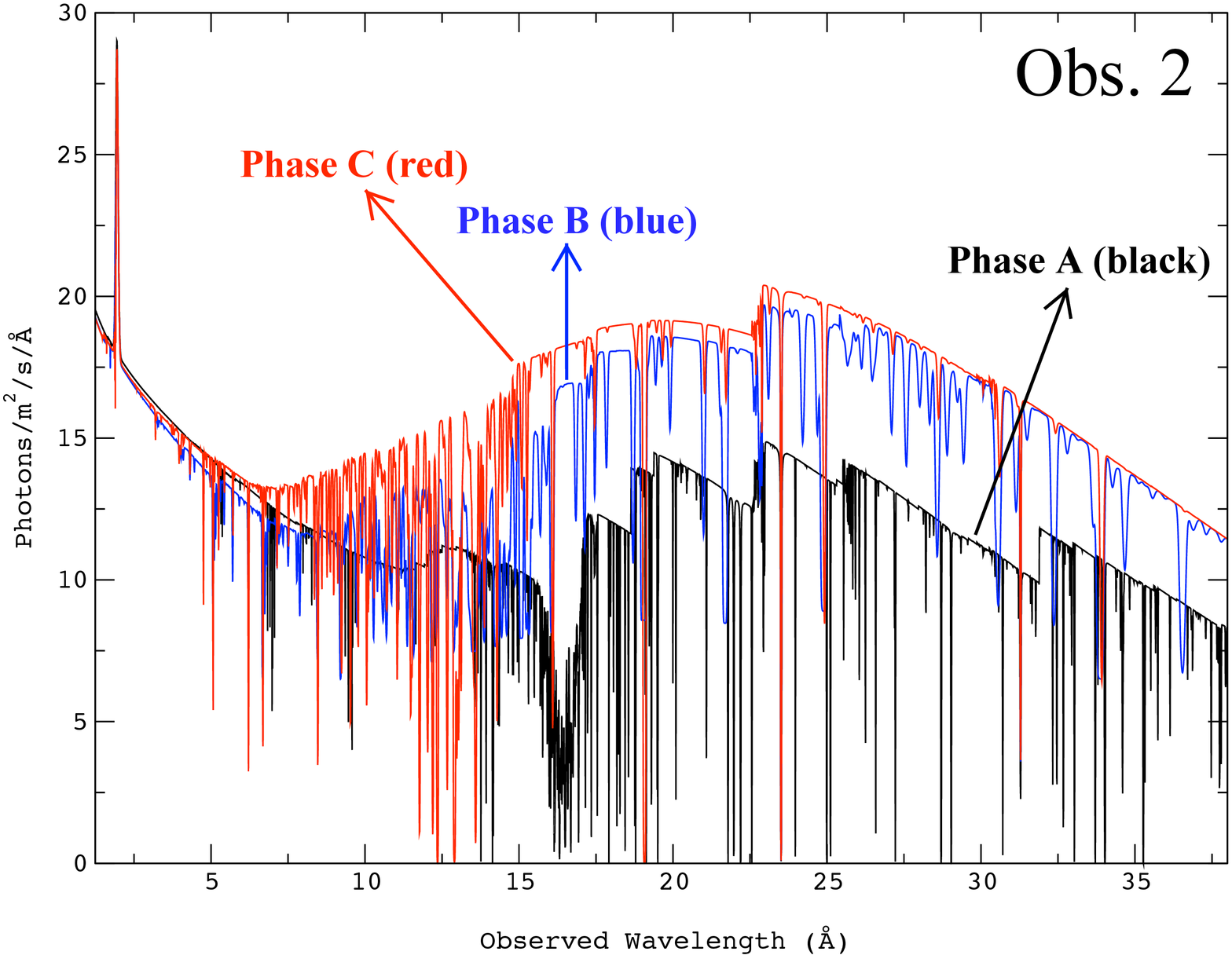}}
\resizebox{\hsize}{!}{\includegraphics[angle=270]{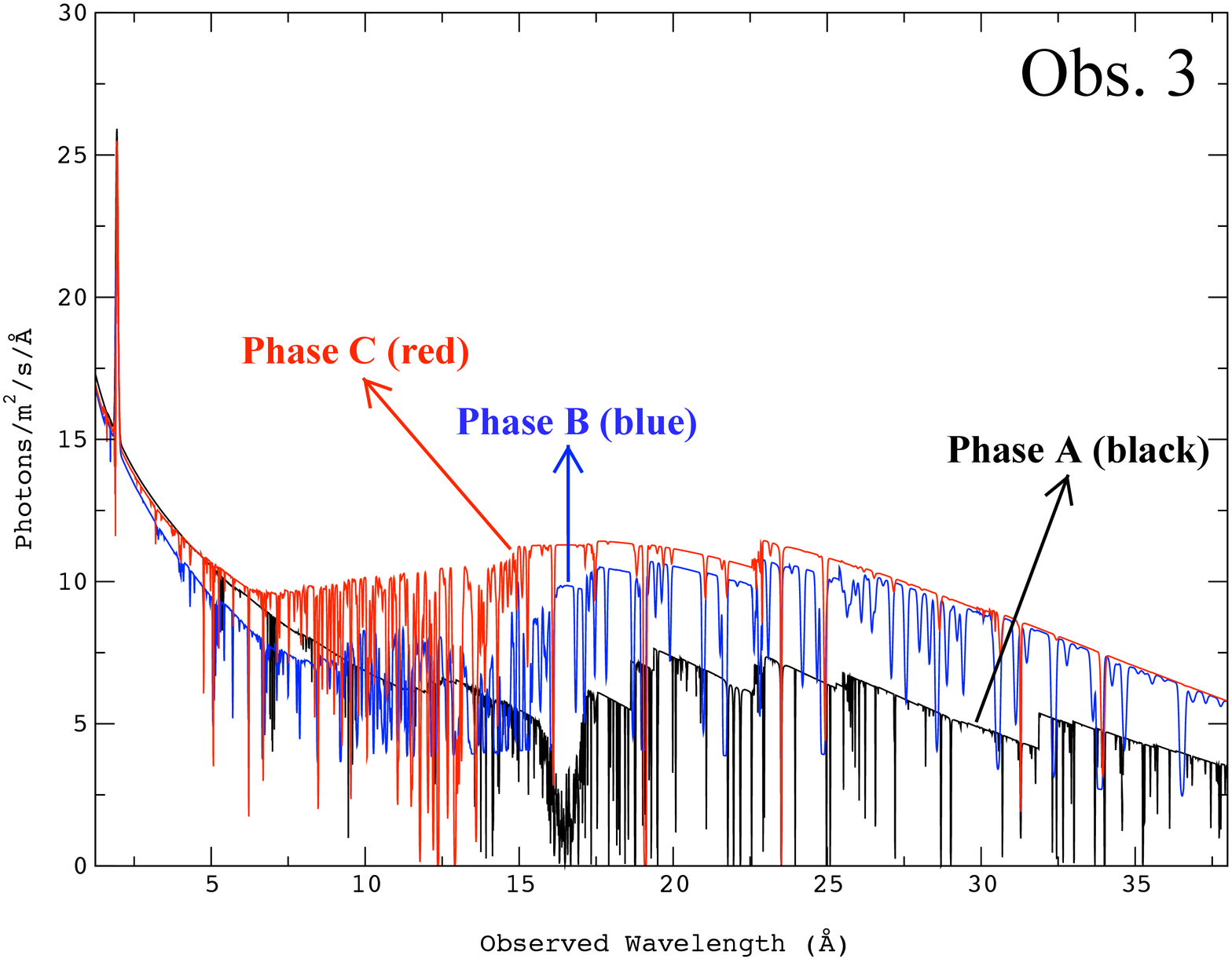}}
\caption{Our best fit warm absorber \xabs model showing the three phases applied separately to the continuum (phase A: black, phase B: blue, phase C: red) from simultaneous fitting of the EPIC-pn and RGS spectra of Obs. 2 (top) and Obs. 3 (bottom). The best fit parameters of the models are given in Tables \ref{cont_table}, \ref{xabs_table} and \ref{abun_table}. The models for Obs. 1 and Obs. 4 look identical to that of Obs. 2 (top).} 
\label{xabs_models}
\end{figure}

Figure \ref{xabs_models} shows how each one of the three phases in our warm absorber model contributes to the best fit for the four observations. Phases A, B and C are plotted in black, blue and red, respectively. Phase A, which is the lowest ionisation phase, is associated with M-shell Fe absorption forming the Unresolved Transition Array (UTA) \citep{Beh01} between 16 and 17 $\AA$. Phase C, the highest ionisation phase, is associated with highly ionised Fe. Phase B, of intermediate ionisation, is the phase with the partial covering fraction.

\begin{figure}
\resizebox{\hsize}{!}{\includegraphics[angle=90]{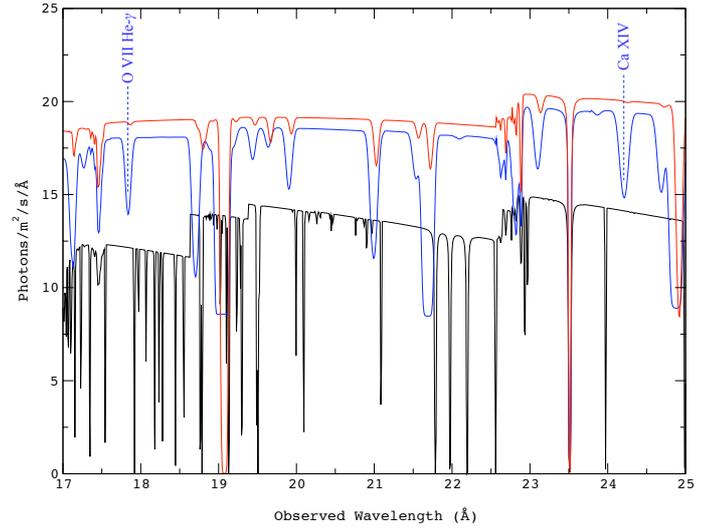}}
\caption{The Obs. 2 best fit \xabs model, showing that the \ion{O}{vii} He-$\gamma$ and \ion{Ca}{xiv} absorption lines are only modelled in phase B. Phase A is shown in black, phase B in blue and phase C in red.}
\label{ovii_caxiv_model}
\end{figure}

In the previous subsection (Sect. \ref{obs3}), we showed that by changing only the covering fraction of phase B, one cannot fit the Obs. 3 spectrum. To confirm that \fcov of phase B does not change significantly between observations, we examined our model closely to look for absorption lines unique to phase B (i.e. that do not appear in phases A and C) and to see if they change between Obs. 2 and Obs. 3. This is important because by observing a change in the depth of these lines, we can investigate the effect of a covering fraction change. We found three of such absorption lines, which are only modelled by phase B. They are \ion{O}{vii} He-$\gamma$ ($17.8\ \AA$), \ion{Ca}{xiv} ($24.1\ \AA$) and \ion{S}{xii} ($36.4\ \AA$). Figure \ref{ovii_caxiv_model} shows the presence of \ion{O}{vii} He-$\gamma$ and \ion{Ca}{xiv} lines in our model; the \ion{S}{xii} line is not displayed separately to save space, but the following is also true for the \ion{S}{xii} line. Comparing the fits in Figs. \ref{ovii_caxiv} (Obs. 2) and \ref{good_fit} (Obs. 3), we can see that the \ion{O}{vii} He-$\gamma$ and \ion{Ca}{xiv} are fitted well in both. It is important to note that in both figures, the covering fraction value is the same (i.e. 0.55). If the covering fraction of phase B had changed significantly in Obs. 3, we would have expected the \ion{O}{vii} He-$\gamma$ and \ion{Ca}{xiv} lines in Fig. \ref{good_fit} not to be fully fitted. The fact that they are fitted well by varying only the continuum parameters is an extra piece of evidence that the variability seen in the lightcurve is mostly intrinsic to the source, and not the warm absorber.

\subsection{Narrow and broad emission features}
\label{emission_sec}
\begin{figure}
\centering
\resizebox{\hsize}{!}{\includegraphics[angle=270]{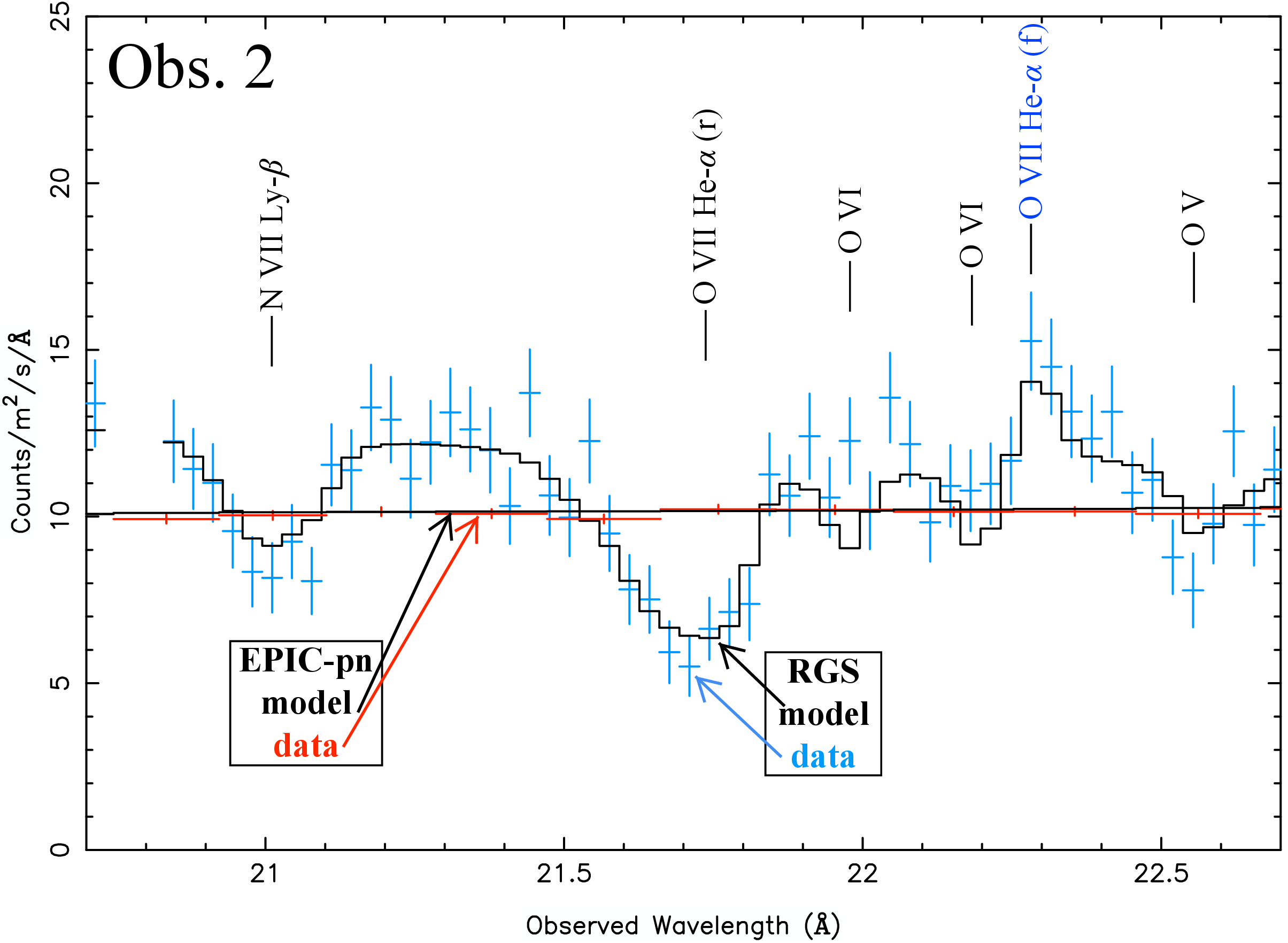}}
\resizebox{\hsize}{!}{\includegraphics[angle=270]{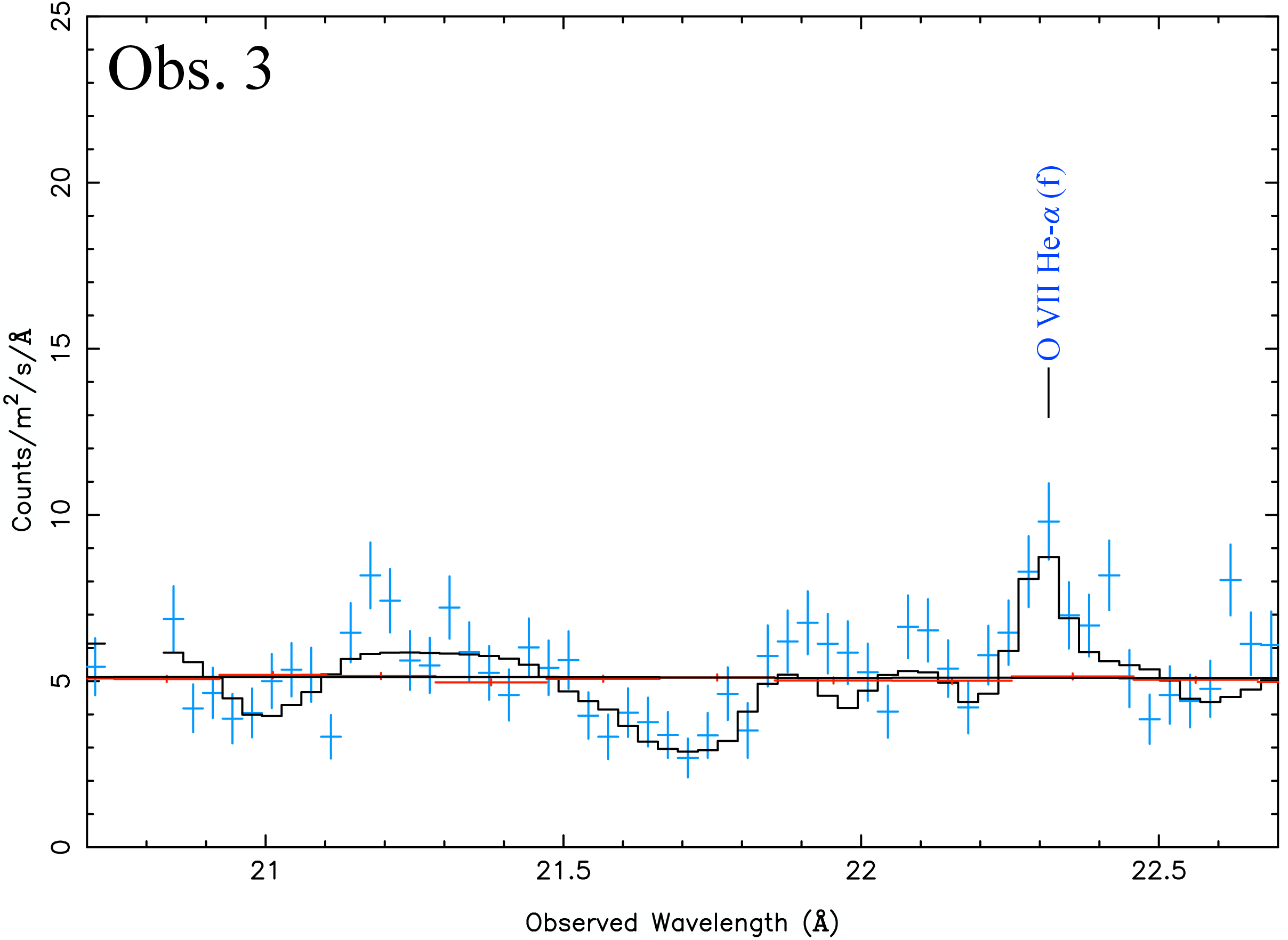}}
\caption{The presence of forbidden \ion{O}{vii} line emission in Obs. 2 (top) and Obs. 3 (bottom). The RGS data are shown in blue, the EPIC-pn data in red and the models in black.}
\label{Oviif}
\end{figure}

The only significant narrow emission feature in the RGS spectra of the four observations of NGC 3516 is the \ion{O}{vii} forbidden line (rest frame wavelength of $22.10\ \AA$). Figure \ref{Oviif} shows the Obs. 2 and Obs. 3 spectra and best fit around the \ion{O}{vii} triplet region. This line was modelled using a Gaussian profile ({\it gaus}) and its best fit parameters are given in Table \ref{emission_table}. The RGS spectrum of NGC 3516 around the \ion{O}{vii} triplet is very similar to that of NGC 5548 presented by \citet{Ste03}: in that case, the forbidden line is present in emission and outflowing at around 110 \kms, the intercombination line is absent or very weak and the resonance line is seen in absorption; there are also absorption lines by \ion{O}{v} and \ion{O}{vi}. As suggested by \citet{Ste03}, one possible explanation for not detecting \ion{O}{vii} intercombination emission is blending of the intercombination line with the \ion{O}{vi} absorption line.

\begin{table*}
\setlength{\extrarowheight}{3pt}
\begin{minipage}[t]{\hsize}
\caption{The best fit parameters of the narrow \ion{O}{vii} forbidden emission line shown in Fig. \ref{Oviif}, obtained by adding a {\it gaus} component to the best fit \xabs model of all four observations. The errors correspond to a $\Delta \chi^{2}$ of 2.}
\label{emission_table}
\centering
\renewcommand{\footnoterule}{}
\begin{tabular}{l l l l l}
\hline \hline
Parameter & Obs. 1 & Obs. 2 & Obs. 3 & Obs. 4 \\
\hline
$\lambda_0$ \footnote{Theoretical rest frame wavelength in \AA.} & $22.10$ & $22.10$ & $22.10$ & $22.10$ \\ 
$\lambda$ \footnote{Wavelength in the rest frame of NGC 3516 in \AA.} & $22.11 \pm 0.06$ & $22.09 \pm 0.04$ & $22.11 \pm 0.02$ & $22.12 \pm 0.08$ \\ 
Flow $v$ \footnote{\kms.} & $+100 \pm 800$  & $-100 \pm 500$ & $+100 \pm 300$ & $+300 \pm 1100$\\
FWHM \footnote{\AA.} & $<0.1$ & $<0.1$ & $<0.1$ & $<0.1$ \\ 
$\sigma_{v}$ \footnote{RMS velocity in \kms ($\sigma_{v}=\mathrm{FWHM}/\sqrt{\ln 256}$).} & $<600$ & $<600$ & $<600$ & $<600$ \\ 
Line Normalisation \footnote{$10^{49}$ photons $\mathrm{s}^{-1}$.} & $1.0 \pm 0.6$ & $1.0 \pm 0.6$ & $1.7 \pm 0.5$ & $0.9 \pm 0.7$ \\ 
\hline
\end{tabular}
\end{minipage}
\end{table*}

\begin{figure}
\centering
\resizebox{\hsize}{!}{\includegraphics[angle=270]{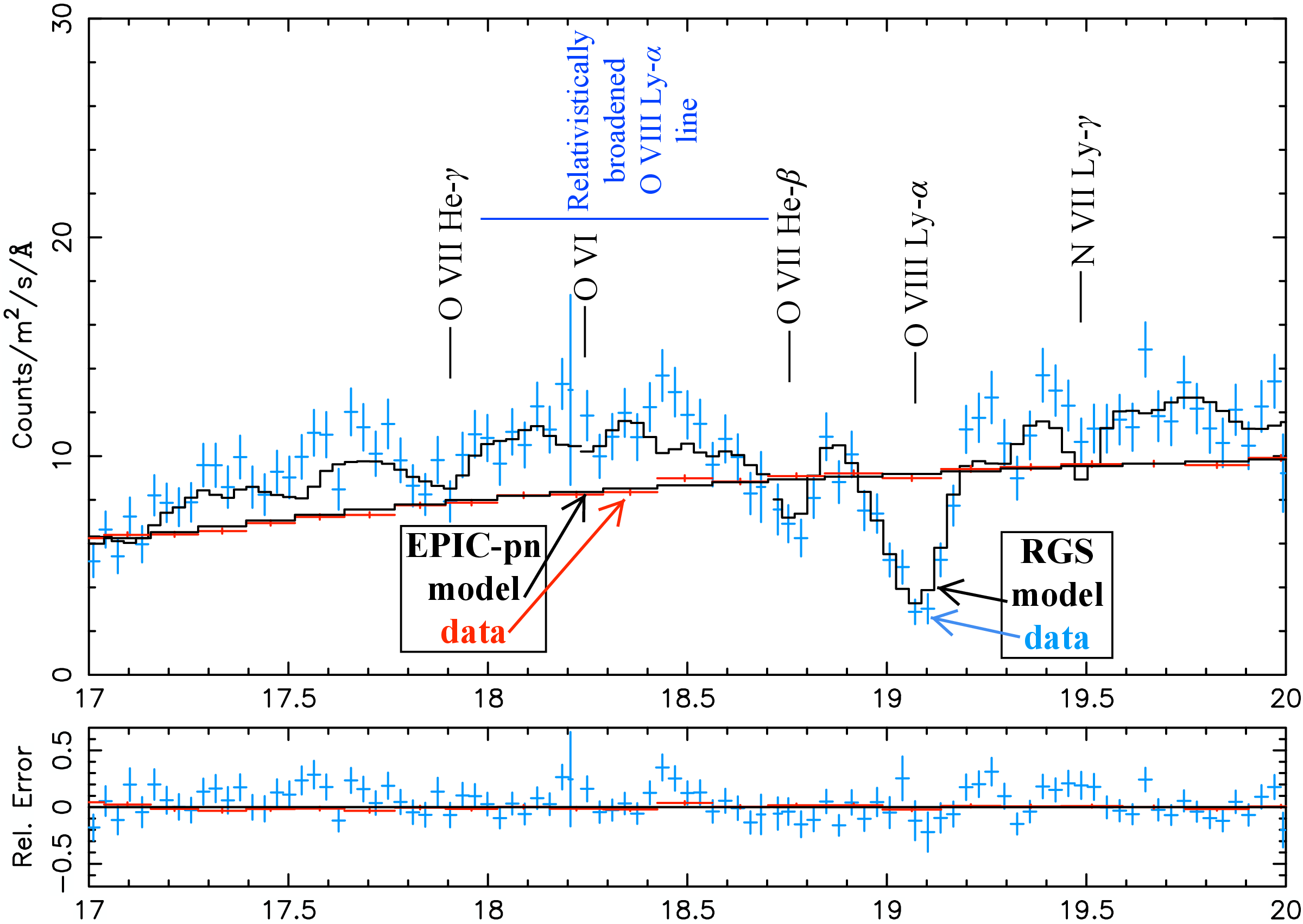}}
\resizebox{\hsize}{!}{\includegraphics[angle=270]{fig12b.ps}}
\caption{Top panel: the Obs. 2 best fit \xabs model fitted simultaneously to the RGS and EPIC-pn spectra with the relativistic emission line included (compare with Fig. \ref{ovii_caxiv} where the line is not included). The RGS data are shown in blue, the EPIC-pn data in red and the models in black. Middle panel: residuals of the fit in the top panel. Bottom panel: the Obs. 2 best fit model of the relativistic \ion{O}{viii} Ly-$\alpha$ emission line added to the continuum. The model is shown in the observed frame, and for clarity of presentation no absorption component of our model is included in this figure.}
\label{rel_model}
\end{figure}

From Figs. \ref{ovii_caxiv} and \ref{good_fit} an excess of emission is noticeable in the RGS data at around 18.0 to 18.5 \AA. This excess is observed in all four observations. Also, in this region of the spectrum, phase A, the lowest-ionised phase, produces absorption by \ion{O}{vi}. We first fitted the excess in Obs. 2 with a simple Gaussian profile; best fit line wavelength of $18.16 \pm 0.04\ \AA$ (in the rest frame of the AGN) and FWHM of $0.47 \pm 0.09\ \AA$ ($\sigma_{v}=3300 \pm 600\ \kms$) were obtained with a \redchi of 1.40 (2781 d.o.f.). The wavelength of the emission line does not correspond to any prominent transition, unless this is emission by \ion{O}{viii} Ly-$\alpha$ (rest-frame $\lambda$ of 18.97 \AA) with an outflow velocity of about 13000 \kms . Therefore, as an alternative we investigated whether the excess could be caused by a relativistically-broadened \ion{O}{viii} accretion disk emission line. If emission from the inner disk is influenced by general relativistic effects, the lines would appear to be broad and asymmetric. Such lines were first proposed to explain the spectra of the Narrow Line Seyfert 1 galaxies \object{MCG $-$6-30-15} and \object{Mrk 766} \citep{BrR01}. We added a narrow emission line ({\it delt} component in SPEX) at the rest frame wavelength of the \ion{O}{viii} Ly-$\alpha$ line (18.97 \AA) at the redshift of NGC 3516, and convolved it with the relativistic disk line profile of \citet{Lao91} ({\it laor}). Parameters of the {\it laor} component are the disk inner and outer radii $r_{\mathrm{in}}$ and $r_{\mathrm{out}}$, its inclination angle $i$ and the emissivity index $q$, which appears in the emissivity law, taken to be proportional to $(r^2  + h^2 )^{ - q/2}$, where $r$ is the radius in the accretion disk and $h$ a scale height \citep{Kaa02}. By adding the relativistic \ion{O}{viii} line to the \xabs best fit model of Obs. 2, our fit improved as \redchi fell from 1.44 (2784 d.o.f.) to 1.40 (2779 d.o.f.) and the excess was well fitted. The line profile and the best fit are shown in Fig. \ref{rel_model}.

%
\begin{table*}
\begin{minipage}[t]{\hsize}
\setlength{\extrarowheight}{3pt}
\caption{Comparison of the ionisation parameters, hydrogen column densities and outflow velocities of the NGC 3516 warm absorber phases found in recent observations, shown in increasing order of $\xi$.}
\label{compare_table}
\centering
\renewcommand{\footnoterule}{}
\begin{tabular}{l l l | l l l | l l l l | l l l l}
\hline
\hline
\multicolumn{3}{c|}{\citet{Tur05} \footnote{From Apr. and Nov. 2001 \xmm EPIC-pn spectra. The values of the ionisation parameter in $\xi$ form are given in \citet{Tur08}.}} & \multicolumn{3}{c|}{\citet{Mar08} \footnote{From Oct. 2005 {\it Suzaku} XIS and HXD spectra.}} & \multicolumn{4}{c|}{\citet{Tur08} \footnote{From Oct. 2006 \xmm EPIC-pn and {\it Chandra} HETG spectra.}} & \multicolumn{4}{c}{Present work \footnote{From Oct. 2006 \xmm EPIC-pn and RGS spectra.}} \\
$\log\ \xi$ \footnote{$\mathrm{erg\ cm\ }\mathrm{s}^{-1}$.} $\sim$ & $\NH$ \footnote{$10^{22}$ $\mathrm{cm}^{-2}$.} $\sim$ & $v$ \footnote{Flow velocity in \kms .} $\sim$ & $\log\ \xi$ $^e$ $\sim$ & $\NH$ $^f$ $\sim$ & $v$ $^g$ $\sim$ & $\log\ \xi$ $^e$ $\sim$ & $\NH$ $^f$ $\sim$ & $v$ $^g$ $\sim$ & Zone & $\log\ \xi$ $^e$ $\sim$ & $\NH$ $^f$ $\sim$ & $v$ $^g$ $\sim$ & Phase \\
\hline
- & - & - & - & - & - & $-2.4$ & $0.2$ & ? & $1$ & - & - & - & -\\
$-0.5$ & $0.5$ & $-200$ & $0.3$ & $5.5$ & ? & $0.25$ & $0.05$ & ? & $2$ & $0.9$ & $0.4$ & $-100$ & A \\
$2.5$  & $25$ & $-1100$ & - & - & - & $2.2$ & $20$ & $-1600$ & $3$ & $2.4$ & 2 & $-1500$ & B \\
$3.0$ & $2$ & $-1100$ & $3.7$ & $4$ & $-1100$ \footnote{Based on the velocity obtained in \cite{Tur05}.} & $4.3$ & $26$ & $-1000$ & $4$ & $3.0$ & $1$ & $-900$ & C \\
\hline
\end{tabular}
\end{minipage}
\end{table*}

The inner disk radius is found to be $14 \pm 3\ GM/c^2$ and the outer disk radius is poorly constrained. The signal of the red tail of the line is not strong enough to be modelled accurately; therefore, as the asymmetry of the line is not very clear in the data, the broadened Gaussian and the Laor profile fit the excess equally well in our modelling; however, as explained earlier, a Gaussian profile does not correspond to any prominent transition. 
 
The disk inclination angle is constrained very well in our model for all four observations; this is because its value very much depends on the position of the bulk of the excess emission in the spectrum: a higher inclination angle causes the emission line profile, shown in Fig. \ref{rel_model} (bottom), to move to lower energies and vice versa. As the excess emission is clear with respect to the continuum, the disk inclination angle is derived to be $33 \degr \pm 2$.

\section{Discussion}
\label{Discussions}

In this work, we have demonstrated the advantage of combining RGS high resolution with EPIC-pn spectroscopy in modelling an AGN warm absorber. EPIC-pn constrains the continuum well, but has insufficient resolution to resolve absorption and emission features seen in the warm absorber (e.g. the Fe UTA and \ion{O}{viii} Ly-$\alpha$ absorption as well as the \ion{O}{vii} He-$\alpha$ (f) emission line shown in Figs. \ref{ovii_caxiv} and \ref{Oviif}). Therefore, to investigate changes in the warm absorber between observations, careful examination and fitting of the RGS spectrum are needed alongside fitting the EPIC-pn spectrum. In the following subsections we discuss our results and their physical implications.

\subsection{The \FeKa line}
In this work we modelled the broad \FeKa line at $\sim$ 6.4 keV with a simple Gaussian profile. For an in-depth spectral analysis of the Fe K regime, we refer the reader to \citet{Tur08} in which the \chandra HETG spectrum is used to model the \FeKa, {Fe K\ensuremath{\beta}\xspace} and \ion{Fe}{xxvi} emission lines in detail. However, from our simple modelling of the line, described in Sect. \ref{continuum} (Table \ref{pow_table}), we find that the \FeKa flux remains practically unchanged between observations, whereas the power-law flux falls significantly in Obs. 3. It is known that the variability of the \FeKa line in general is not correlated to that of the observed continuum in a trivial manner \citep{Min04}. The iron line does not always respond to variations in the continuum: in some cases the \FeKa line can appear to be constant while the continuum varies by a large amplitude (e.g. \citealt{Mar03}). Using the EPIC-pn spectra, we cannot examine the \FeKa profile so accurately as with the {\it Chandra} HETG, so we will not discuss this any further.

\subsection{The warm absorber structure}
\label{abs_discu}
We have found that three phases of ionisation are adequate to model the warm absorption in NGC 3516. The elemental abundances are unlikely to be different in the three phases, since it is possible to fit the spectrum very well by coupling all the abundances at values close to solar (Table \ref{abun_table}). There is also no need for a large over-abundance of iron to fit the UTA.

A comparison of $\xi$, \NH and outflow velocity values of warm absorber phases from recent observations of NGC 3516 reported by different authors is shown in Table \ref{compare_table}. The ionisation parameters of our phases B and C are remarkably similar to those found by \cite{Tur05} during the 2001 observations. We do not find requirement for absorption by a fourth phase (identified by {\it Chandra} HETG to correspond to $\log\ \xi \sim 4.3$ in \citealt{Tur08}) from our simultaneous EPIC-pn and RGS fittings, which are not very sensitive to such high ionisation gas. Note that $\log\ \xi$ values are not strictly comparable as they are SED dependent, although the differences are likely to be only of the order of some tenth of dex. The outflow velocity of our partially covering phase B is practically the same as the one in \cite{Tur08} (their partially covering Zone 3); these phases also have very similar ionisation parameters, suggesting they represent the same photoionised gas. The second phase in \cite{Tur05}, which was also partially covering, is again very similar to our phase B in terms of ionisation parameter and outflow velocity (albeit this is smaller in 2005). The pattern that emerges by comparing the outflow velocities of all the phases found by different authors is the existence of low-ionised phases with very small outflow velocity and higher ionisation phases with outflow velocities between $1000$ \kms and $1500$ \kms.

The partially covering phase is the phase with the highest outflow velocity and most likely closest to the central engine. Since we do not find changes in the occultation of the nuclear source, there is no evidence of a transverse component of velocity for the phase. The minimum distance of this phase from the central engine can be estimated assuming the outflow velocity we measure to be greater than or equal to the escape velocity $v_{{\rm{esc}}}  = \sqrt {2GM/r}$, where $G$ is the gravitational constant, $M$ the black hole mass and $r$ the distance of the absorber phase from the black hole. Since the partially covering phase B is outflowing at $1500$ \kms, using the NGC 3516 black hole mass estimate of $2.95 \times 10^7\ M_{\odot}$ \citep{Nik06}, we obtain $r \gtrsim 0.1\ \mathrm{pc}$. Using the expression for the ionisation parameter definition (Eq. \ref{ion_eq}) and taking $L_{\mathrm{ion}} \simeq 3.85 \times 10^{43}\ \mathrm{erg}\ \mathrm{s}^{-1}$ (derived from the fit of Obs. 2) we find the hydrogen number density $n \lesssim 1.3 \times 10^8\ \mathrm{cm}^{-3}$. Combining this with the relation $\NH \sim n \Delta r$, where $\Delta r$ is the thickness of the absorber phase, we find $\Delta r \gtrsim 5 \times 10^{-5}\ \mathrm{pc}$. This suggests the presence of a thin spherical shell of gas, which is a feasible scenario for partial covering.

The column densities \NH of all three phases of our warm absorber cover a relatively small range: $\sim$ 0.4--2 $\times 10^{22}\ \mathrm{cm}^{-2}$, and we do not find ``heavy" ($\gtrsim 20 \times 10^{22}$ $\mathrm{cm}^{-2}$) absorber phases as in \citet{Tur05,Tur08}, especially those corresponding to our phase B. However, our range of \NH is consistent with the values obtained from other X-ray observations of NGC 3516: \citet{Cos00} found a warm absorber column of $\sim$ 2--3 $\times 10^{22}\ \mathrm{cm}^{-2}$ from {\it BeppoSAX} spectra; \citet{Net02} found a line-of-sight absorber with \NH of $\sim 1 \times 10^{22}\ \mathrm{cm}^{-2}$ from both a 1994 {\it ASCA} observation and a 2000 {\it Chandra} observation; \citet{Mar08} found their two warm absorber phases to have \NH of $\sim$ 4 and 5.5 $\times 10^{22}\ \mathrm{cm}^{-2}$.

\citet{Tur05} concluded that their three absorber phases cannot be in thermal equilibrium with each other in NGC 3516. We can consider the photoionisation equilibrium curve technique (S-curve, \citealt{Kro81}) for the SED used in our analysis (NGC 5548, \citealt{Ste05}). The S-curve marks the points of temperature $T$ versus pressure form of the ionisation parameter $\Xi$ at which the pressure $P$ is constant. The parameter $\Xi$ is related to $\xi$ by
\begin{equation}
\Xi  = \frac{L_{\mathrm{ion}}}{{4\pi cr^2 P}} = 0.961 \times 10^4 \frac{\xi }{T}
\end{equation}
where $L_{\mathrm{ion}}$ is the luminosity of the ionising source, $c$ is the speed of light, $r$ is the distance of the warm absorber phase from the ionising source, $P$ is the pressure and $T$ the temperature. \citet{Ste05} found that the S-curve for NGC 5548 does not have negative slope anywhere; applying this to NGC 3516, we can conclude that the three absorption phases we identify in NGC 3516 outflowing with different velocities are unlikely to be in pressure equilibrium. However, it is worth noting that different photoionisation codes, variations in the shape of the SED and selecting different energy ranges for the construction of the SED (\citealt{Smi09} and \citealt{Cha09}) can produce some differences in the S-curve. Because the ionisation parameters, column densities and outflow velocities of our phases B and C are similar (although not in pressure equilibrium), and more importantly the covering fraction of phase B is not variable, phase B could have the same origin as the fully covering phase C. This disfavours a clumpy disk wind scenario as the most likely explanation for the origin of phase B.

\subsection{Narrow and broad emission features}

The RGS spectrum shows evidence for both narrow and broad emission features. The \ion{O}{vii} (f) emission line is found to have a lower flow velocity compared to the observed absorption lines of the high ionisation phases B and C; the flow velocity is in fact consistent with zero, and with that of the lowest ionisation phase A. \citet{Smi08} have observed the \ion{O}{vii} (f) line to have a very low outflow velocity compared to other soft X-ray emission lines in \object{Ark 564}, although in that case all absorption phases also have flow velocities consistent with zero. This lack of a velocity shift from the rest wavelength is in agreement with a scenario in which we are observing emission from a fully visible spherically symmetric outflowing shell around the nuclear source, thus no net velocity of \ion{O}{vii} will be detected. Another possible explanation is that the \ion{O}{vii} line originates in the Narrow Line Region (NLR), where the emission lines are produced in a gas of comparatively small velocity dispersion and low density (see for e.g. \cite{Kaa02}, where the forbidden \ion{O}{vii} line in NGC 5548 is suggested to come from the NLR). As the forbidden line is the most intense line in the \ion{O}{vii} triplet, a photoionised gas with low density is required.

We have found possible evidence for a relativistically broadened emission line of \ion{O}{viii} Ly-$\alpha$. However, the line is weaker than the strong \ion{O}{viii}, \ion{N}{vii} and \ion{C}{vi} Ly-$\alpha$ lines reported by \citet{BrR01} and \citet{Sak03} in MCG $-$6-30-15 and Mrk 766. Evidence for weak broadened emission lines of \ion{O}{viii} and \ion{N}{vii} Ly-$\alpha$ is also found in NGC 5548 by \citet{Kaa02}. The disk inclination angle obtained from our fit to the \ion{O}{viii} Ly-$\alpha$ profile is $33 \degr \pm 2$. \citet{WuH01, Zha02} have calculated the inclination angles of the Broad Line Region (BLR) of several Seyfert 1 galaxies using already published bulge stellar velocity dispersions and black hole masses estimated by reverberation mapping. They calculated an inclination angle of $38.3 \degr \pm 7.6$ for NGC 3516, which would imply co-planarity of the BLR with the AGN accretion disk.

\subsection{Intrinsic continuum versus warm absorber variability}
NGC 3516 has a history of large amplitude continuum variability between observations. \citet{Net02} reported a large drop in flux (factor of $\sim 50$ at 1 keV) between an {\it ASCA} observation in 1994 and a {\it Chandra} observation in 2000. They found that the observed flux and spectral variability at these epochs were consistent with a constant column density of line-of-sight material reacting to changes in the ionising continuum. In the 2006 \xmm data we have found the observed spectral and flux variability to be unrelated to changes in the covering fraction of phase B, unlike what is reported by \citet{Tur08}. Furthermore, the X-ray absorption line depths are sensitive to changes in the covering fraction; from close examination of RGS high resolution spectra, there is no evidence to suggest that the covering fraction of phase B changed between observations.

The only parameters which indicate a change during the low flux \xmm Obs. 3 are the ionisation parameter $\xi$ of phase A and the column density $\NH$ of phase B. We find $\log\ \xi \sim 0.9$ and $\NH \sim 3 \times 10^{22}\ \mathrm{cm}^{-2}$ in Obs. 3, whereas in the other three observations they are $\sim 1.0$ and $\sim$ 2 $\times 10^{22}\ \mathrm{cm}^{-2}$, respectively. However, these changes in the warm absorber parameters are very small compared to those in the continuum parameters: power-law slope $\Gamma$ from $\sim 1.8$ to 1.7 in Obs. 3, and normalisation from $\sim 3$ to $2 \times 10^{51}$ photons $\mathrm{s}^{-1}$ $\mathrm{keV}^{-1}$ at 1 keV; modified black body temperature from $\sim 190$ to 210 eV in Obs. 3, and normalisation (emitting area times square root of electron density) from $\sim 1.5$ to $0.5 \times 10^{33}$ $\mathrm{cm}^{0.5}$.

The power-law contribution to the continuum heavily outweighs the modified black body in all our four observations and is responsible for nearly all of the continuum variability. From the X-ray lightcurve (Fig. \ref{PN_lightcurve}) and spectrum (Fig. \ref{PN_all}), the variation in Obs. 3 is larger in the soft X-ray (0.2--2.0 keV) energy band than in the hard X-ray (2.0--10.0 keV) band. This type of variability has been seen in other X-ray observations of Seyfert 1 AGN (such as \object{NGC 7469}, \citealt{Blu03}) in which the source is softer when brighter.
\section{Conclusions}
\label{Conclusions}
\begin{enumerate}

\item We have studied the warm absorber in NGC 3516 by analysing in detail EPIC-pn and RGS high resolution spectra from four observations made in October 2006 by \xmm. The warm absorber consists of three phases of ionisation: phase A ($\log\ \xi \sim 0.9$), phase B ($\log\ \xi \sim 2.4$) and phase C ($\log\ \xi \sim 3.0$) in increasing order of ionisation. Phase A has a hydrogen column density of $\sim 0.4 \times 10^{22}\ \mathrm{cm}^{-2}$, which is smaller than those of the other two phases ($\sim 1-2 \times 10^{22}\ \mathrm{cm}^{-2}$). There is evidence that the lower-ionisation phase A is outflowing at $\sim 100\ \kms$, whereas the two higher-ionisation phases are outflowing faster at around 1000 to 1500 \kms.

\item Phase B covers about 60\% of the source continuum. We investigated whether variation in the covering fraction of phase B could account for the observed flux and spectral variability in Obs. 3 (as claimed by \citealt{Tur08}). We found that: (1) the covering fraction does not show significant variation between observations; (2) even if the covering fraction is altered significantly, this does not properly account for the observed variability in the EPIC-pn and especially RGS spectra. This makes a clumpy disk wind scenario a rather unfeasible explanation.

\item Similar to \citet{Net02}, we conclude that the variably in the 2006 observations presented here (albeit much smaller than in previous observations) is better understood as the consequence of changes in the source continuum emission than in the warm absorber. Our results suggest that the X-ray variability of NGC 3516 (and by inference, possibly that of other AGN where a similar behaviour has been observed) cannot be reduced to occultation and absorption effects, such as proposed by \citet{Tur08} (see also review by \citealt{Tur09}, and references therein); rather, the variability is likely to arise in a scenario where intrinsic changes of the continuum and ionisation state are also important. In this context, a careful analysis of the soft X-ray spectrum at high resolution, in combination with data in the Fe Ka regime, can provide essential constraints and clues to the source physical behaviour.

\end{enumerate}

%
\begin{acknowledgements}
This work is based on observations obtained with \xmm, an ESA science mission with instruments and contributions directly funded by ESA member states and the USA (NASA). Missagh Mehdipour acknowledges the support of a PhD studentship awarded by the UK Science \& Technology Facilities Council (STFC). We thank the referee, Elisa Costantini, for all her useful comments that improved the paper.
\end{acknowledgements}

%
%

\end{document}